\documentclass[final,3p,times]{elsarticle}
\usepackage{amssymb}
\usepackage{amsmath}
\journal{Physica A}

\def\Eq#1{Eq.~(\ref{#1})}
\def\Eqs#1{Eqs.~(\ref{#1})}
\def\Fig#1{Fig.~\ref{#1}}	
\def\nd{\noindent}
\def\no{\nonumber\\}
\def\r{\rangle\!\rangle}
\def\l{\langle\!\langle}
\def\>{\rangle}
\def\<{\langle}
\def\u#1{\underline{#1\!}\,	}

\def\adg{a^\dagger}
\def\tr{\text{tr}}
\def\dg{\dagger}
\def\xt{\tilde{x}}
\def\lam{\lambda}

\def\om{\omega}
\def\wt{\tilde{\omega}}
\def\sig{\sigma}

\def\del{\delta}
\def\gam{\gamma}

\def\bt{\beta}

\def\d{\partial}

\def\KL{\text{KL}}
\def\Kt{\tilde{K}}
\def\Kb{\bar{K}}

\def\CL{\text{CL}}
\def\m{\text{mKL}}
\def\HPZ{\text{HPZ}}
\def\KB{\boldsymbol{K}}
\def\TB{\boldsymbol{T}}
\def\FB{\boldsymbol{F}}

\def\Qt{\tilde{Q}}
\def\rt{\tilde{r}}
\def\Qb{\bar{Q}}
\def\rb{\bar{r}}

\def\Ft{\tilde{F}}
\def\Fm#1{F_N^{(#1)}}
\def\F#1#2{F_{#1}^{(#2)}}
\def\Fb#1#2{\bar{F}_{#1}^{(#2)}}
\def\G#1#2{G_{#1}^{(#2)}}
\def\H#1#2{H_{#1}^{(#2)}}

\def\Ph#1#2{\Phi_{#1}^{(#2)}}
\def\Phib#1#2{\bar{\Phi}_{#1}^{(#2)}}
\def\Phit#1#2{\tilde{\Phi}_{#1}^{(#2)}}
\def\phiB{\boldsymbol{\phi}}

\def\u#1#2{u_{#1}^{(#2)}}
\def\v#1#2{v_{#1}^{(#2)}}
\def\w#1#2{w_{#1}^{(#2)}}
\def\p#1#2{p_{#1}^{(#2)}}
\def\q#1#2{q_{#1}^{(#2)}}
\def\s#1#2{s_{#1}^{(#2)}}
\def\x#1#2{x_{#1}^{(#2)}}

\def\fb{\bar{f}}
\def\ft{\tilde{f}}
\def\bb{\bar{b}}
\def\bt{\tilde{b}}
\def\Pib{\bar{\Pi}}
\def\Omb{\bar{\Omega}}

\def\Psib#1#2{\bar{\Psi}_{#1}^{(#2)}}
\def\Psit#1#2{\tilde{\Psi}_{#1}^{(#2)}}

\begin{document}

\begin{frontmatter}

\title{Liouvillian exceptional points in continuous variable system}

\author[1]{B. A. Tay}
\ead{BuangAnn.Tay@nottingham.edu.my}
\address[1]{Department of Foundation in Engineering, Faculty of Science and Engineering, University of Nottingham Malaysia, Jalan Broga, 43500 Semenyih, Selangor, Malaysia}

\date{12 April 2023}

\begin{abstract}
The Liouvillian exceptional points for a quantum Markovian master equation of an oscillator in a generic environment are obtained. They occur at the points when the modified frequency of the oscillator vanishes, whereby the eigenvalues of the Liouvillian become real. In a generic system there are two parameters that modify the oscillator's natural frequency. One of the parameters can be the damping rate. The exceptional point then corresponds to critical damping of the oscillator. This situation is illustrated by the Caldeira--Leggett (CL) equation and the Markovian limit of the Hu--Paz--Zhang (HPZ) equation. The other parameter changes the oscillator's effective mass whereby the exceptional point is reached in the limit of extremely heavy oscillator. This situation is illustrated by a modified form of the Kossakowski--Lindblad (KL) equation. The eigenfunctions coalesce at the exceptional points and break into subspaces labelled by a natural number $N$. In each of the $N$-subspace, there is a $(N+1)$-fold degeneracy and the Liouvillian has a Jordan block structure of order-$(N+1)$. We obtain the explicit form of the generalized eigenvectors for a few Liouvillians.
Because of the degeneracies, there is a freedom of choice in the generalized eigenfunctions. This freedom manifests itself as an invariance in the Jordan block structure under a similarity transformation whose form is obtained.
We compare the relaxation of the first excited state of an oscillator in the underdamped region, critically damped region which corresponds to the exceptional point, and overdamped region using the generalized eigenvectors of the CL equation.
\end{abstract}


\end{frontmatter}

\section{Introduction}

In open quantum systems, the generators governing the evolution of a system are
intrinsically dissipative and hence non-Hermitian. At certain choices of
parameters of the system, the eigenvalues of the generator and the corresponding
eigenvectors could coalesce. The points in the parameter space at which the
eigenvalues coalesce are called exceptional points~\cite{Heiss12,Berry04}. The
exceptional points are different from degeneracy of eigenvalues in conservative
systems. Whereas the generators of evolution in conservative systems can be
diagonalized at degenerate eigenvalues, those in dissipative systems can at
best be brought into a Jordan canonical form. In scattering theory, the
coalescence of eigenvalues manifests itself as double or higher order poles in
the complex energy plane~\cite{Mondragon93,Bhamathi96,Bohm97,Ferise22} that
lead to resonances.

The occurrence of exceptional points are ubiquitous in dissipative systems.
It was observed in microwave cavity~\cite{Heiss01}, optical
cavity~\cite{Lee09}, atom--cavity~\cite{Choi10},
electronics~\cite{Schindler19}, photonics~\cite{Ozdemir19,Quiroz-Juarez19},
driven superconducting qubits~\cite{Chen21} and etc.
It was found that at the exceptional points, the enhanced gain or loss of the
system, or a balance between them could potentially lead to novel applications.
This had been exploited to create enhanced sensors~\cite{Wiersig20,Yuan22},
non-reciprocal topological energy transfer~\cite{Xu16}, revival of
lasing~\cite{Peng14}, and etc.
There were also connections between exceptional points and the symmetry
breaking in PT symmetric Hamiltonians~\cite{Guo09} as well as the onset of
quantum phase transitions~\cite{Heiss05}.

Recently, there are increasing interests in extending the understanding of exceptional points physics from the level of Hamiltonian to the level of Liouvillian, i.e.,~the evolution generator of quantum master equation.
The Liouvillian of quantum master equation is a non-Hermitian
superoperator~\cite{Petrosky97}. Along this line, the exceptional points of the
Liouvillians of finite-level systems were explored in Ref.~\cite{Hatano19} and
from the perspective of quantum jumps in Refs.~\cite{Nori19,Nori20b,Nori20c}.

In this work we focus on the Liouvillian of an infinite-level or continuous variable system.
Utilizing the results obtained recently on the spectrum of the Liouvillian of a
quantum oscillator in a generic environment~\cite{Tay20}, we obtain the
exceptional points of the Liouvillian at arbitrary order and the generalized
eigenfunctions from the Jordan block structure of the Liouvilian.

We first discuss the degeneracy in the Kossakowski--Lindblad (KL) equation in  \ref{SecEigevK}. Then, the approach to exceptional points in a generic quantum master equation are elucidated with two examples in  \ref{SecEP}. The generalized eigenvectors are obtained in  \ref{SecGJME}. This is followed by a discussion on the invariance of the Jordan block structure of the reduced dynamics under similarity transformation in  \ref{SecGJvec}. We illustrate the results by comparing the evolution of a quantum state in three regions of damping in  \ref{SecEvoEP}. A conclusion then follows. Some of the details of the calculations are presented in the appendices.

\section{Degeneracy in  {Kossakowski--Lindblad} ({ {KL}}) equation}
\label{SecEigevK}

We consider a quantum Markovian master equation, $\d \rho/\d t=-K\rho$, for the density operator $\rho$ of a harmonic oscillator. The effect of the environment is encoded in the Liouvillian $K$.
We start by considering the Kossakowski--Lindblad (KL)
equation~\cite{Kossa76,Lindblad76} (also known as the
Gorini--Kossakowski--Sudarshan--Lindblad or simply the Lindblad equation) which is
widely used to describe the influence of environment on an oscillator in many
fields~\cite{Nielsen,Gardiner,Breuer,May11}, such as in quantum information,
quantum optics, condensed matter, energy transfer in molecular systems, and
etc.
Its Liouvillian is usually written in terms of the creation and annihilation operators of a harmonic oscillator as
\begin{equation}
\label{KKLaa}
K_\KL\rho=i\om_0 [\adg a,\rho]
    -\frac{\gam}{2}\left(b+\frac{1}{2}\right)(2a\rho\adg-\adg a\rho-\rho\adg a)
    -\frac{\gam}{2}\left(b-\frac{1}{2}\right)(2\adg\rho a-a\adg\rho-\rho a\adg )\,,
\end{equation}
 where $\gam$ is a relaxation rate and $b$ is related to the temperature $T$ of the environment by $b\equiv \frac{1}{2}\coth[\hbar\om/(2kT)]$.
At absolute zero when $b=1/2$, the last group of operators in $K_\KL\rho$ drops out.

To continue our discussion, we specialize in the position coordinates.
We first introduce the dimensionless position coordinate $x\equiv \sqrt{m\om_0/\hbar}\,q$, where $q$ is the position coordinate with a dimension of length. In the Liouville space, we have the bra-space $\lvert \xt \r$ and the ket-space $\l x \rvert $. The density function in the position coordinates is denoted by $\langle x| \rho| \xt\rangle$.
It is customary to introduce the centre and relative coordinates
\begin{equation}
 \label{Qrxx}
    Q \equiv\frac{1}{2}(x+\xt)\,,   \qquad r\equiv x-\xt\,,
\end{equation}
 respectively, and write the density function as
\begin{equation}
 \label{rhoQr}
    \langle x| \rho| \xt \rangle =\left\langle Q+\frac{r}{2}\bigg| \rho\bigg| Q-\frac{r}{2}\right \rangle \equiv\rho(Q,r)\,.
\end{equation}
 The annihilation operator in the position coordinates is $a=(\hat{x}+i\hat{p})/\sqrt{2}=(\hat{x}+\d/\d \hat{x})\sqrt{2}$, where $\hat{x}$ and $\hat{p}=-i\d/\d\hat{x}$ are the dimensionless position and momentum operators.
Then,
\begin{equation}
 \label{arhox}
    \langle x| a\rho| \xt \rangle =\frac{1}{\sqrt{2}}\left(x+\frac{\d}{\d x}\right)\langle x| \rho| \xt \rangle
    =\frac{1}{\sqrt{2}}\left(Q+\frac{r}{2}+\frac{1}{2}\frac{\d}{\d Q}+\frac{\d}{\d r}\right)\rho(Q,r)\,.
\end{equation}
 Similarly,
\begin{equation}
 \label{rhoax}
    \langle x| \rho a| \xt \rangle
    =\frac{1}{\sqrt{2}}\left(\xt-\frac{\d}{\d \xt}\right)\langle x| \rho | \xt \rangle
    =\frac{1}{\sqrt{2}}\left(Q-\frac{r}{2}-\frac{1}{2}\frac{\d}{\d Q}+\frac{\d}{\d r}\right)\rho(Q,r)\,.
\end{equation}
\Eqs{arhox} and \eqref{rhoax} are defined under a trace, see \ref{SecNormGV}.
In the process of getting \Eq{rhoax}, an integration by parts is carried out with an assumption that $\langle x| \rho| \xt \rangle $ vanishes fast enough at the infinity. $\langle x| \rho| \xt \rangle $ with the form of a Gaussian multiplying a polynomial is a useful example of functions satisfying this requirement.
The corresponding expressions of creation operator in the position coordinates can be obtained similarly by using $\adg=(\hat{x}-i\hat{p})/\sqrt{2}=(\hat{x}-\d/\d \hat{x})\sqrt{2}$.

In this way, in the position coordinates the Liouvillian $K_\KL$ can be simplified to
\begin{equation}
\label{KKL}
K_\KL(Q,r)=i\om \left(-\frac{\d^2}{\d Q\d r}+Qr\right)
    - \frac{\gam}{2}\left(\frac{\d}{\d
    Q}Q-r\frac{\d}{\d r}\right)
    - b \frac{\gam}{2}\left(\frac{\d^2}{\d Q^2}-r^2\right)\,.
\end{equation}
 The eigenvalue equation
\begin{equation}
\label{Keig}
K_\KL f^\pm_{mn}=\lam^\pm_{mn}f^\pm_{mn}\,, \qquad 0\leq n\leq m=0,1,2, \ldots\,,
\end{equation}
 has the solution~\cite{Briegel93,Tay08,Honda10}
\begin{equation}
\label{lampm}
\lam^\pm_{mn}=\pm in\om +(2m-n)\frac{\gam}{2}\,,\qquad 0\leq n\leq m=0,1,2, \ldots\,.
\end{equation}
 The eigenfunctions $f^\pm_{mn}(Q,r)$ can be decomposed into two parts~\cite{Tay08},
\begin{equation}
   \label{fQr}
    f^\pm_{mn}(Q,r)
    = \Pi^\pm_{mn}(Q,r) f_{00}(Q,r)\,,\qquad 0\leq n\leq m=0,1,2, \ldots\,,
\end{equation}
 where
\begin{equation}
  \label{f00}
            f_{00}(Q,r)=\frac{1}{\sqrt{2 \pi b}}
           \,  e^{-Q^2/(2b)-br^2/2}
\end{equation}
 is a stationary state, whereas $\Pi^\pm_{mn}(Q,r)$ is a polynomial in $Q$ and $r$. Its expression is given in \ref{AppPi} for the convenience of the reader. In the following discussion, we will omit the coordinates dependence on the functions when confusion does not arise.

\begin{figure}[t]
\centering
\includegraphics[width=4.2in, trim = 4cm 10cm 7cm 11cm]{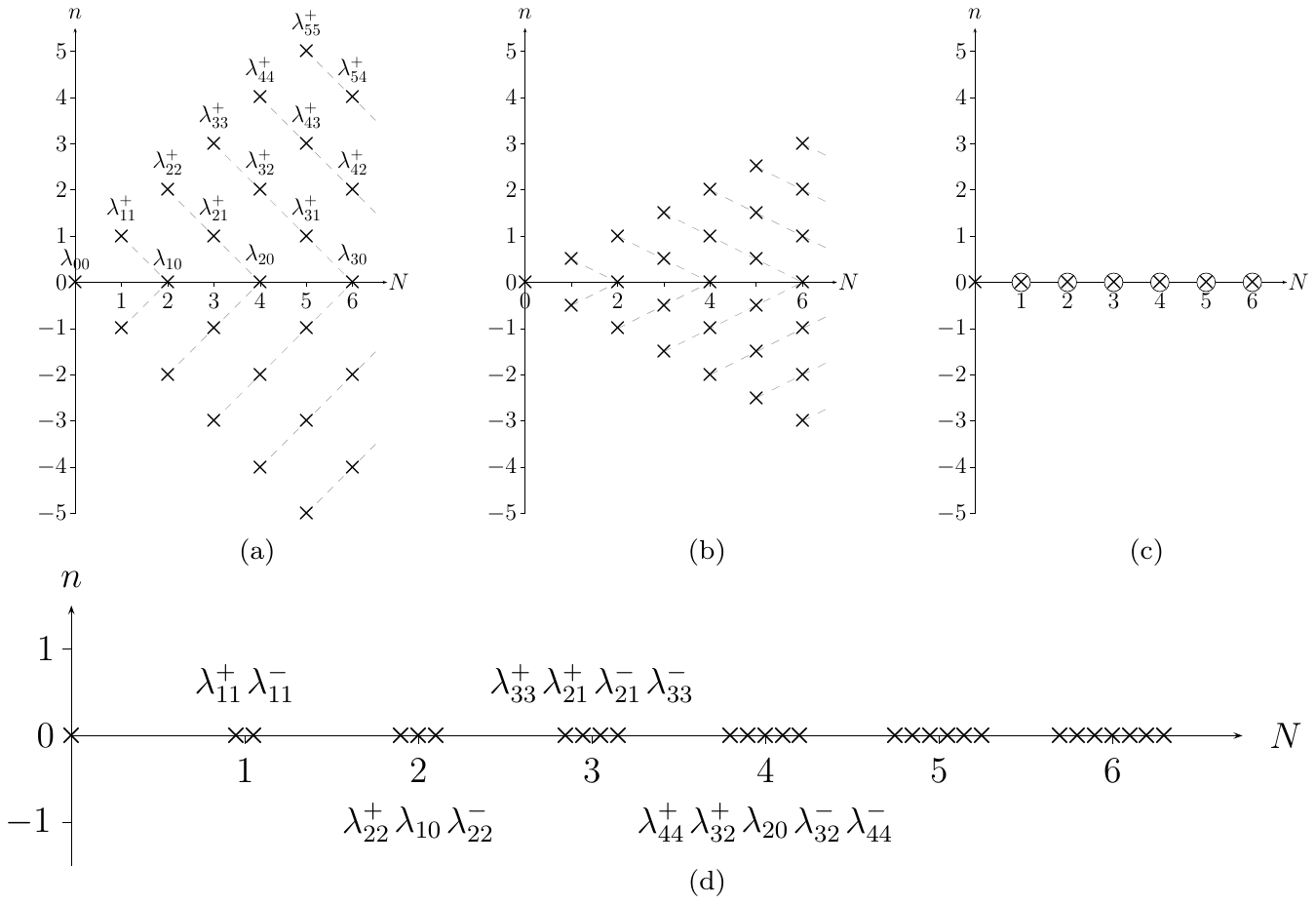}
\caption{The change in the positions of eigenvalues $\lam_{mn}^\pm$ \eqref{lampm} on a complex plane as real $\om$ reduces to 0 (the exceptional point) and then turns imaginary.
Horizontal and vertical axis are scaled in units of $\gam/2$ and $\om$, respectively.
(a) The eigenvalues away from the exceptional point. Dashed lines connect eigenvalues with the same $m=(N+n)/2$. The corresponding eigenvalue is given above the cross. The position of $\lam_{mn}^-$ is a mirror reflection of $\lam_{mn}^+$ along the horizontal axis.
(b) As $\om$ decreases, the eigenvalues move vertically towards the horizontal axis from both the above and below of the axis.
(c) At the exceptional point, the eigenvalues $\lam_N$ \eqref{lamN} coalesce on the horizontal axis and become real. They are labelled by circles with a cross. They have a $(N+1)$-fold degeneracy.
(d) When $\om$ turns imaginary, the degeneracies are removed. The eigenvalues split away from each other along the horizontal axis. The distance between two nearest eigenvalues of the same $N$-subspace is $2| \om| $. The eigenvalues that correspond to the crosses are given in the respective order up to $N=4$ for illustration.\label{fig1}}
\end{figure}

\Fig{fig1}(a) gives the positions of the eigenvalues \eqref{lampm} on a complex plane.
As $\om$ decrease to zero, two or more eigenvalues such as $\lam^\pm_{mn}$ and $\lam^\pm_{m'n'}$, are degenerate whenever the condition $2m'-n'=2m-n$ is met.
Both the $\lam^+_{mn}$ and $\lam^-_{mn}$ move vertically towards the real horizontal axis, see \Fig{fig1}(b), and meet there at $\om=0$, see \Fig{fig1}(c), at which the eigenvalues become real.
At this point, it is convenient to introduce the label
\begin{equation}
   \label{N}
    N\equiv 2m-n\,, \qquad N=0,1,2,3, \ldots
\end{equation}
 to write the eigenvalue as
\begin{equation}
   \label{lamN}
    \lam_N\equiv N\frac{\gam}{2}\,.
\end{equation}
 Each $N$ form a subspace of eigenvectors with a $(N+1)$-fold degeneracy.
Note that for even $N$, there is already an existing eigenvalue $\lam_{N/2,0}$ on the real axis.
When $\om$ becomes imaginary, the eigenvalues remain real.
The degeneracies that occur at $\om=0$ are lifted and the eigenvalues split from each other along the horizontal axis, see \Fig{fig1}(d). The separation between two nearest eigenvalues of the same $N$-series is $2| \om| $.
In analogy to a damped oscillator in classical mechanics \cite{Symon}, \Fig{fig1}(a) and \Fig{fig1}(b) represent the case of an underdamped oscillator, \Fig{fig1}(c) a critically damped oscillator, and \Fig{fig1}(d) an overdamped oscillator.

For the KL equation, the eigenfunctions of degenerate eigenvectors remain distinct.
As a result, the structure of the eigenvalue equation remains intact.
This is because the unitary part of the Liouvillian $K_0$ (the terms on the right-hand-side (RHS) of \Eq{KKL} containing $\om$), and the rest of its dissipative part $K_{\text{d}}$ containing $\gam$ are separable, i.e.,~they separately satisfy the eigenvalue equation
\begin{equation}
\label{K0Kd}
K_0f^\pm_{mn}=\pm in\om f^\pm_{mn}\,, \qquad
    K_{\text{d}}f^\pm_{mn}=(2m-n)\frac{\gam}{2} f^\pm_{mn}\,.
\end{equation}
 Consequently, the dynamics of the unitary and dissipative part do not affect each other in the limit of zero $\om$.

We illustrate this with the $m=1$ subspace.
Taking the limit $\om\rightarrow0$, there are two degenerate eigenvalues $\lam^+_{11}=\lam^-_{11}=\gam/2$.
The eigenfunctions remain linearly independent,
\begin{equation}
\label{f11w=0}
f^\pm_{11}=\mp i\left(\frac{Q}{\sqrt{2b}}\pm \sqrt{\frac{b}{2}}r\right)f_{00}
\end{equation}
 Consequently, there is a 2-fold degeneracy in this subspace
\begin{equation}
\label{f11}
K_\KL f^\pm_{11}=\frac{\gam}{2}f^\pm_{11}\,.
\end{equation}
 We note that $f^+_{11}$ and $f^-_{11}$ remain orthogonal to each
other~\cite{Tay08}.

\section{Exceptional points in quantum master equation}
\label{SecEP}

In previous works~\cite{Talkner81,Tay17}, it was shown that the Liouvillian of
a generic quantum Markovian master equation has the form
\begin{equation}
   \label{Kgen}
    K
    =2\om_0 iL_0+h_1iM_1+h_2iM_2+\gam(O_0-I/2)+g_0O_++g_1L_{1+}+g_2L_{2+}\,.
\end{equation}
The $h_i$ and $g_i$ are constant parameters, whereas $\gam$ is a damping or relaxation constant of the system.
The expressions of the seven operators in the position coordinates are given in \ref{AppPi}.
We note that $K_\KL$ \eqref{KKLaa} is obtained by choosing $g_0=-2\gam b$ and zero for the rest of the parameters.
$K$ can be related to the Liouvillian of the KL equation by a
similarity transformation $K=SK_\KL S^{-1}$~\cite{Tay20}, provided that the modified
frequency of system,
\begin{equation}
\label{w0h}
\om=\sqrt{\om_0^2-\frac{h_1^2}{4}-\frac{h_2^2}{4}}\,,
\end{equation}
 satisfies the condition
\begin{equation}
   \label{whcond}
    \om^2+\frac{\gam^2}{4}\neq0\,.
\end{equation}
 Then, the eigenvalue of the transformed system is also given by \Eq{lampm}, with eigenfunction $Sf^\pm_{mn}$.

In elementary mechanics, there are three regions of harmonic motion of an
oscillator under a friction force~\cite{Symon,Heiss16}, i.e.,~underdamping,
critical damping and overdamping that correspond to real $\om>0$,
$\om=0$ and $\om$ imaginary, respectively.
The consideration carries over into the reduced dynamics of a quantum oscillator.
From the discussion in the previous section, degeneracy occurs when $\om=0$ at critical damping. For a given natural frequency $\om_0$, points of degeneracy lie on a circle of radius $2\om_0$ in the $(h_1, h_2)$-parameter space
\begin{equation}
   \label{degen}
    h_1^2+h_2^2=4\om_0^2\,.
\end{equation}

We will consider two independent ways in the approach to $\om=0$, (1) from $h_2\rightarrow 2\om_0$ with $h_1=0$ in the Caldeira--Leggett (CL) equation, and (2) from $h_1\rightarrow2\om_0$ with $h_2=0$ in the modified Kossakowski--Lindblad (KL) equation.
For generic quantum master equation, the approach to zero $\om$ exhibits behaviour that is distinct from the KL equation.

\subsection{ {Caldeira--Leggett} ({ {CL}}) equation}
\label{SecEigevCL}

We first consider the Caldeira--Leggett (CL) equation~\cite{CL83} with
$h_2=-\gam$, $g_0=g_1=-2\gam \bar{b}$ in \Eq{Kgen}, whereas the rest of the parameters are zero.
We will used ``bar'' to denote quantities related to the CL equation.
In terms of the creation and annihilation operators of harmonic oscillator, the CL equation is written as
\begin{equation}
\label{KCLaa}
\Kb_\CL=i\om_0[\adg a,\rho]+i\frac{\gam}{2}[\hat{x},\{\hat{p},\rho\}]+\gam \bb[\hat{x},[\hat{x},\rho]]\,,
\end{equation}
where $\{\cdot,\cdot\}$ denotes an anti-commutator bracket.
In the high temperature limit, the second term with the anti-commutator bracket
can be dropped, as is usually done in the studies on environmental
decoherence~\cite{Decoh03,Dodd04Halliwell}.

In the position coordinates, the Liouvillian of the CL equation is
\begin{equation}
\label{KCLQr}
\Kb_\CL(\Qb,\rb)=i\om_0\left(-\frac{1}{2}\frac{\d^2}{\d\Qb\d\rb}+2\Qb\rb\right)
        +\gam\rb\frac{\d}{\d\rb}+2\gam\rb^2\,,
\end{equation}
where $\bar{b}$ is absorbed into the coordinates
\begin{equation}
   \label{Qrbar}
  \Qb\equiv\frac{Q}{\sqrt{2\bb}}\,,\qquad \rb\equiv\sqrt{\frac{\bb}{2}}r\,.
\end{equation}

$\Kb_\CL$ is similarly related to $K_\KL$ by $\Kb_\CL=S_2 K_\KL S_2^{-1}$~\cite{Tay20},
where
\begin{equation}
   \label{SCL}
    S_2=e^{i\phi M_1} e^{\eta L_{2+}}\,.
\end{equation}
 The stationary state is
\begin{equation}
   \label{f00CL}
  \fb_{00}(\Qb,\rb)=S_2 f_{00}(Q,r)=\frac{1}{\sqrt{\pi}}e^{-\Qb^2-\rb^2} \,.
\end{equation}
 The corresponding function $\bar{\Pi}^\pm_{mn}$ in \Eq{fQr} can be obtain in a similar
way from $f^\pm_{mn}$. We refer the reader to Ref.~\cite{Tay20} for the
details.
A few $\fb^\pm_{mn}$ are listed in \ref{AppEPdamp}.

The damped oscillator has the modified frequency
\begin{equation}
 \label{w2}
    \om_2\equiv\sqrt{\om_0^2-\gam^2/4}  \,.
\end{equation}
 For the approach of $\om_2$ to zero at critical damping, we consider the limit $\gam\rightarrow2\om_0$ from the side $\gam<2\om_0$ so that $\om_2$ remains real.
In this limit, the eigenvalues are still given by \Eq{lamN}. However, small denominator could develop in the eigenfunctions leading to divergence.
As an example, we illustrate this with the eigenfunction $\fb^\pm_{11}$ under the $N=1$ subspace of the CL equation.
In \ref{AppEPdamp}, we expand $\fb^\pm_{11}$ in powers of the  parameter
\begin{equation}
   \label{CLdel}
    \del_2 \equiv 1-\frac{\gam}{2\om_0}\,.
\end{equation}
 In the limit $\om_0\rightarrow\gam/2$, the most singular term in the series is
\begin{equation}
\label{Fb1pm}
\bar{f}_{11}^{\pm}=\frac{1}{\sqrt{\del_2}}\frac{1\mp i}{2}(\Qb-i\rb)\fb_{00}+O(\del_2^0)\,,
\end{equation}
 which has a square root singularity.
Eigenfunctions with higher order $N$ will have increasingly higher power of square root singularity.
The singularity shows that this is not a case of simple degeneracy.
Furthermore, notice that both eigenvectors are the same up to an overall constant, i.e.,~they are no longer orthogonal.
$\om_2=0$ is called the exceptional point to distinguish it from ordinary degeneracy.

Another Liouvillian with a similar behaviour is the Markovian limit of the HPZ
equation~\cite{HPZ92,Tay07} which is related to the CL equation by a similarity
transformation.
Owing to its similarity with the CL equation, it approaches the exceptional point in a similar way with the CL equation.
We present the results for the HPZ equation in \ref{AppHPZ}.

\subsection{Modified  {Kossakowski--Lindblad} ({ {KL}}) equation}
\label{SecEigevKmod}

We consider another model with a different approach to the exceptional points.
We add a $h_1iM_1$ term \eqref{iM1} to $K_\KL$ \eqref{KKLaa} to get a modified KL equation. The Liouvillian becomes
\begin{equation}
\label{KLmQr}
\Kt_\m(\Qt,\rt)=-\frac{i}{2}\left(\om_0-\frac{h_1}{2}\right)\frac{\d^2}{\d\Qt\d\rt}
    +2i\left(\om_0+\frac{h_1}{2}\right)\Qt\rt
    -\frac{\gam}{2}\left(\frac{\d}{\d\Qt}\Qt-\rt\frac{\d}{\d\rt}
    +\frac{1}{2}\frac{\d^2}{\d \Qt^2}-2\rt^2\right)\,,
\end{equation}
 where we define
\begin{equation}
   \label{Qrtilde}
  \Qt\equiv\frac{Q}{\sqrt{2\bt}}\,,\qquad \rt\equiv\sqrt{\frac{\bt}{2}}r\,,
\end{equation}
and use ``tilde'' to denote quantities related to the modified KL equation.

Due to the interaction between the oscillator and the environment, the oscillator acquires an effective mass
\begin{equation}
   \label{effm}
    \tilde{m} \equiv m \left(1-\frac{h_1}{2\om_0}\right)^{-1}\,.
\end{equation}
The frequency of the oscillator is modified to
\begin{equation}
 \label{w1}
    \om_1\equiv\sqrt{\om_0^2-h^2_1/4}\,.
\end{equation}
We approach the limit of heavy oscillator $h_1\rightarrow2\om_0$ from the side $h_1<2\om_0$, so that $\om_1$ is real.

$\Kt$ is similarly related to $K_\KL$ through
\begin{equation}
\label{S1}
S_1= e^{\eta_0 L_{1+}}e^{\psi iM_2}e^{\eta_2 iL_{2+}}\,.
\end{equation}
Using the method discussed in Ref.~\cite{Tay20}, we apply the transformation
to the stationary state of the KL equation to get the stationary state of
$\Kt_\m$,
\begin{equation}
\label{f00KLmod}
\tilde{f}_{00}(\Qt,\rt)=S_1f_{00}=\frac{1}{\sqrt{\pi}} \sqrt{\frac{\om_1^2+\gam^2/4}{\om_0(\om_0-h_1/2)+\gam^2/4}}
  \exp\left[-
  \frac{(\om_1^2+\gam^2/4)\Qt^2+\gam h_1 i\Qt\rt/2+(\om_0^2+\gam^2/4)\rt^2
       }{\om_0(\om_0-h_1/2)+\gam^2/4}
       \right]\,.
\end{equation}
 The corresponding transformed polynomial $\tilde{\Pi}^\pm_{mn}=S_1\Pi^\pm_{mn}S_1^{-1}$, cf.~\Eq{fQr}, can be worked out in the same way.
A few of them can be found in \ref{AppEPKLmod}.

In the case of the modified KL equation, we show in \ref{AppEPKLmod} that in the limit $h_1\rightarrow2\om_0$, two of the eigenfunctions in the $N=1$ subspace go into
\begin{equation}
\label{Ft1p}
\tilde{f}_{11}^\pm=-\sqrt{2}\wt ir\tilde{f}_{00}
    +O\left(\sqrt{\del_1}\right)\,,
\end{equation}
where
\begin{equation}
   \label{wt}
    \wt \equiv \frac{2\om_0}{\gam}\,.
\end{equation}
In contrast to the eigenfunctions of $\Kb_\CL$ which develop small denominator, the eigenfunctions of $\Kt_\m$ contain increasing power of the square root of a small numerator
\begin{equation}
   \label{del1}
        \del_1\equiv 1-\frac{h_1}{2\om_0}\,.
\end{equation}
Furthermore, both the $\tilde{f}^+_{11}$ and $\tilde{f}^-_{11}$ are identical in this limit.
This again implies that we lose a copy of linearly independent vector.
Hence, $\om_1=0$ is the exceptional point for the modified KL equation.

It is known in scattering theory that whenever the $S$-matrix has a
pole of order $N$, its Hamiltonian when extended beyond the Hilbert
space reveals a Jordan block structure of order
$N$~\cite{Bhamathi96,Bohm97}.
On the level of density operator, we have seen that a change of parameters causes the eigenvalues to coalesce.
At the exceptional points, the usual eigenvalue equation is no longer capable of describing the reduced dynamics.
The eigenvalue equation has to be generalized to a Jordan block structure,
\begin{subequations}
\begin{align}
  \label{KGJ0}
    K_E\Fm{0}&=\lam_N\Fm{0}\,,\\
    K_E\Fm{z}&=\lam_N\left(\Fm{z}+\Fm{z-1}\right)\,,\qquad z=1,2, \ldots,N\,,  \label{KGJz}
\end{align}
\end{subequations}
 for an exceptional point of order $N+1$.
In the following section we will construct higher-order generalized eigenvectors by solving the Jordan block structure \eqref{KGJ0}--\eqref{KGJz} in a systematic way.
There, we realize that \Eqs{Fb1pm} and \eqref{Ft1p} give the zeroth-order vector $\Fm{0}$ up to a normalization constant.

We note that in the literature, the coefficient of $\Fm{z-1}$ in \Eq{KGJz} is usually chosen as 1.
The discrepancy is due to a different definition in the normalization constant of $\Fm{z}$.
The definition we adopt render all $\Fm{z}$ dimensionless.
We will discuss the normalization of the generalized eigenfunctions in \ref{SecNormGV}.

\section{Generalized eigenvectors of quantum master equation}
\label{SecGJME}

In this section we construct the generalized eigenvectors of the Liouvillian of CL and modified KL at the exceptional points.
Similar to the eigenvectors of $K_\KL$ \eqref{fQr}, we decompose the generalized eigenvectors into two parts,
\begin{equation}
\label{GJdecomp}
\F{N}{z}=\Ph{N}{z} \F{0}{0}\,,
\end{equation}
 where $\F{0}{0}$ is the stationary state and $\Ph{0}{0}=1$.
$\Ph{N}{z}$ is a polynomial in the coordinates $Q$ and $ir$ with real coefficients. In \Eq{GJdecomp} and the equations to follow, we omit the coordinates dependence on the functions to simplify the expressions.

Before we construct the eigenvectors, we discuss the hermiticity requirement on the density operator $\rho^\dg=\rho$. In the coordinate representation, this requirement is translated into $\langle x| \rho^\dg| \xt \rangle =\langle \xt| \rho| x \rangle ^{*}=\langle x| \rho| \xt \rangle $, or equivalently, $\rho^{*}(Q,-r)=\rho(Q,r)$.
In the language of superoperator, a superoperator with this property is also
called adjoint-symmetric~\cite{Prigogine73,Tay17}.
We can put a function in an
explicitly adjoint-symmetric form by writing the function as a polynomial in
powers of $Q$ and $ir$.
The coefficients in the polynomials are then real numbers.

We find that the $\Ph{N}{z}$ can be decomposed into a combination of lower order generalized eigenvectors multiplying terms that are linear in the coordinates. We start with two ansatz, which we call the diagonal and parallel representation of the generalized eigenvector,
\begin{subequations}
\begin{align}
   \label{dia}
    \text{Diagonal}:\qquad &\Ph{N}{z}=(\u{N}{z} Q+\v{N}{z} ir)\Ph{N-1}{z}+\w{N}{z}\Ph{N-2}{z} \,,\qquad z\neq N\,,\\
         \text{Parallel}:\qquad &\Ph{N}{z}=(\p{N}{z} Q+\q{N}{z} ir)\Ph{N-1}{z-1}+\s{N}{z}\Ph{N-2}{z-2}\,,\qquad z\neq 0\,,\label{par}
\end{align}
\end{subequations}
 where $u,v,w,p,q,s$ are real constant.
By definition, we set $\Ph{N}{z}=0$ whenever $z<0$ or $z>N$ or $N<0$.
We note that additional terms, such as $\x{N}{z}\Ph{N-2}{z-1}$ might need to be added to \Eqs{dia}--\eqref{par} for different choices of generalized eigenvectors, see \ref{AppGJCL} for the details.

We summarize the procedure here. The details are left to \ref{AppGJCL} and \ref{AppGJKL}.m
We start with the subspace $N=1$ by applying \Eq{KGJ0} to the lowest order generalized eigenvector $\F{1}{0}$ using the diagonal representation \eqref{dia}.
By comparing the coefficients of the polynomial in $Q$ and $ir$ on both sides of the equation, we solve for their coefficients.
We next apply \Eq{KGJz} to $\F{1}{1}$ using the parallel representation and solve for the coefficients.
Then we move on to the next subspace of generalized eigenvectors with $N=2$.
In this way we can determine the coefficients of the polynomials from one order to the next for all $N$.

\subsection{ {Caldeira--Leggett} ({ {CL}}) equation}
\label{SecGJCL}

The exceptional points for $\Kb_\CL$ is reached by increasing the damping to $\gam=2\om_0$ where the modified frequency $\om_2$ vanishes.
At the exceptional point, $\Kb_\CL$ \eqref{KCLQr} turns into
\begin{equation}
   \label{KLEPQr}
    \Kb_E(\Qb,\rb) \equiv i\frac{\gam}{2}\left(-\frac{1}{2}\frac{\d^2}{\d\Qb\d\rb}+2\Qb\rb\right)
        +\gam\rb\frac{\d}{\d\rb}+2\gam\rb^2\,.
\end{equation}
The details in obtaining the coefficients in \Eqs{dia}--\eqref{par} are worked out in \ref{AppGJCL}.
The lowest order generalized eigenvector $\Fb{0}{0}$ is identical with $\fb_{00}$ \eqref{f00CL}, and $\Phib{0}{0}=1$.
The higher order generalized eigenvectors can be summarized as
\begin{subequations}
\begin{align}
   \label{PiCLdia}
    \text{Diagonal}:\qquad &\frac{\Phib{N}{z}}{N^z}=(\Qb-i\rb)\frac{\Phib{N-1}{z}}{(N-1)^z}
    -\frac{N-1-z}{2}\frac{\Phib{N-2}{z}}{(N-2)^z}\,,\qquad z\neq N\,,\\
    \text{Parallel}:\qquad &\frac{\Phib{N}{z}}{N^z}=\frac{1}{z}(-i\rb)
    \frac{\Phib{N-1}{z-1}}{(N-1)^{z-1}}\,,\qquad\qquad\qquad\qquad\qquad z\neq 0\,,  \label{PiCLpar}
\end{align}
\end{subequations}
Recall that we set $\Phib{N}{z}=0$ whenever $z<0$ or $z>N$ or $N<0$.
A few examples are listed in
\ref{tabPibC1}.
\begin{table}[t]    \label{tabPibC1}
\begin{center}
\begin{tabular}{c|ccccccccc}
   $z$ &\quad  &$N=0$ &\quad  &$N=1$ &\quad &$N=2$  &\quad &$N=3$   \\
   \hline\\
   0 &\quad  &1 &\quad &$\Qb-i\rb$ &\quad  &$(\Qb-i\rb)^2-\frac{1}{2}$ &\quad &$(\Qb-i\rb)^3-\frac{3}{2}(\Qb-i\rb)$  \\\\
   1 &\quad  & &\quad &$-i\rb$ &\quad &$-2i\rb(\Qb-i\rb)$ &\quad &$-3i\rb\left((\Qb-i\rb)^2-\frac{1}{2}\right)$ \\\\
   2 &\quad  & &\quad  & &\quad &$2(i\rb)^2$
     &\quad &$\frac{9}{2}(i\rb)^2(\Qb-i\rb)$ \\\\
   3 &\quad  & &\quad  &   &\quad  & &\quad  &$-\frac{9}{2}(i\rb)^3$
\end{tabular}
\end{center}
\caption{The first few series of $\Phib{N}{z}$.}
\end{table}
In the overlapping region of the two equations, i.e.,~when $z=1,2,3, \ldots, N-1$, we can show that the two sets of equation \eqref{PiCLdia} and \eqref{PiCLpar} are identical using a proof by induction.
The details are presented in \ref{AppGJCLequal} and \ref{AppGJCLproof}.

This set of generalized eigenvectors is not unique. The non-uniqueness arises from the degeneracies.
It reveals itself as a freedom in the choice of the coefficients in the diagonal and parallel representation.
Examples are given in \ref{AppGJCLdiapar}.
We will elaborate this freedom of choice in  \ref{SecFreedomGJ}.

\subsection{Modified  {Kossakowski--Lindblad} ({ {KL}}) equation}
\label{SecGJKLMod}

In the second approach to the exceptional point in $\Kt_\m$, we increase the effective mass of the oscillator until the modified frequency $\om_1$ vanishes in the limit $h_1\rightarrow 2\om_0$.
In this limit, the Liouvillian $\Kt_\m$ \eqref{KLmQr} can be cast into the form
\begin{equation}
\label{KLmEPQr}
\Kt_E(\Qt,\rt)=4i\om_0\Qt\rt-\frac{\gam}{2}\left(\frac{1}{2}\frac{\d^2}{\d \Qt^2}+\Qt\frac{\d}{\d\Qt}-\rt\frac{\d}{\d\rt}-2\rt^2+1\right)\,.
\end{equation}
The zeroth-order generalized eigenvector $\Ft_{0}^{0}$ is obtained from the stationary state in \Eq{f00KLmod} by setting $h_1=2\om_0$. It can be simplified to
\begin{equation}
\label{F0KLm}
\Ft_0^{(0)}=\frac{1}{\sqrt{\pi}} e^{-\Qt^2-2\wt i\Qt\rt-\left(1+\wt^2\right)\rt^2}\,,
\end{equation}
where $\wt$ is already defined in \Eq{wt}.

In \ref{AppGJKL}, we construct the higher order generalized eigenvectors in increasing  the procedure described in the paragraph after \Eqs{dia}--\eqref{par}.
The results are
\begin{subequations}
\begin{align}
   \label{PiKLmdia}
    \text{Diagonal}:\qquad &\frac{\Phit{N}{z}}{N^z}= i\wt\rt\frac{\Phit{N-1}{z}}{(N-1)^z}\,,\qquad\qquad\qquad\qquad\quad z\neq N\,,\\
    \text{Parallel}:\qquad &\frac{\Phit{N}{z}}{N^z}
        =\frac{\Qt}{2z}\frac{\Phit{N-1}{z-1}}{(N-1)^{z-1}}
        -\frac{1}{8z}\frac{\Phit{N-2}{z-2}}{(N-2)^{z-2}}\,,\qquad z\neq 0\,.  \label{PiKLmpar}
\end{align}
\end{subequations}
 Again, we set $\Phit{N}{z}=0$ whenever $z<0$ or $z>N$ or $N<0$.
The generalized eigenvectors from the first few subspaces are given in
\ref{tabKlmod}.
\begin{table}[t]    \label{tabKlmod}
\begin{center}
\begin{tabular}{c|cccccccccc}
   $z$ &\quad  &$N=0$ &\quad  &$N=1$ &\quad &$N=2$  &\quad &$N=3$   \\
   \hline\\
    0 &\quad &1 &\quad &$i\wt \rt$   &\quad  &$(i\wt\rt)^2$   &\quad    &$(i\wt\rb)^3$\\\\
    1  &\quad & &\quad &$\Qt/2$    &\quad &$i\wt\rt\Qt$ &\quad &$\frac{3}{2}(i\wt\rt)^2\Qt$ \\\\
    2  &\quad & &\quad &          &\quad &$\frac{1}{2}\Qt^2-\frac{1}{4}$ &\quad &$\frac{9}{8}(i\wt\rt)\left(\Qt^2-\frac{1}{2}\right)$\\\\
    3  &\quad & &\quad &     &\quad &  &\quad  &$\frac{9}{16}\Qt\left(\Qt^2-\frac{3}{2}\right)$
\end{tabular}
\caption{The first few series of $\Phit{N}{z}$.}
\end{center}
\end{table}

We can also prove by induction that in the overlapping region of the equations, $z=1,2, \ldots,N-1$, both equations are identical.
The set of generalized eigenvectors again are not unique.
Its origin will be discussed in the next section.

\section{Freedom in the basis of generalized eigenvectors}
\label{SecGJvec}

\subsection{Normalization of generalized eigenvectors}
\label{SecNormGV}

Let us discuss the normalization of generalized eigenvectors.
The trace of a density operator in the coordinate representation is defined as (by setting $\xt=x$ and integrate over $x$)
\begin{equation}
   \label{traceF}
    \tr\rho=\int_{-\infty}^{\infty} \rho(x,r=0) \,dx=1\,.
\end{equation}
 We note that except for the stationary state $F_{0}^{(0)}$, all decaying
vectors have zero trace~\cite{Tay08,Mingati18},
\begin{equation}
   \label{IntF}
    \int_{-\infty}^{\infty}\Fm{z}(x,r=0)\,dx=\del_{N0}\del_{z0}\,.
\end{equation}
 From \Eqs{KGJ0}--\eqref{KGJz}, we deduce that once we fix the overall constant of $\Fm{0}$, the overall constant of higher order vectors in the same series are also fixed.

We mentioned at the end of  \ref{SecEP} that in the literature the coefficient of $\Fm{z-1}$ in \Eq{KGJz}  chosen as 1.
We can bring it into the form of \Eq{KGJz}, or vice versa, by a redefinition of the overall constant in $\Fm{z}$.
In the system we consider, $\lam_N$ is real and non-zero.
We can redefine $\Fm{z}$ as $F_N^{\prime(z)}\equiv \Fm{z}/\lam_N^z$.
Then \Eq{KGJ0} remains unchanged, whereas \Eq{KGJz} becomes
\begin{equation}
   \label{KGJz'}
    KF_N^{\prime(z)}=\lam_N F_N^{\prime(z)}+F_N^{\prime(z-1)}\,,
\end{equation}
 which is the usual way of writing the generalized eigenvalue equation in the literature.
By adopting the normalization implied by \Eqs{KGJ0}--\eqref{KGJz}, all the $\Fm{z}$ are dimensionless.
We can also simplify the expressions when we discuss the invariance of the Liouvillain at the exceptional point in the next section.

\subsection{Freedom in generalized eigenvectors}
\label{SecFreedomGJ}

As already mentioned in the last section, the set of generalized eigenvectors we have constructed are not unique because of degeneracies.
Equally valid sets of generalized eigenvectors satisfying \Eqs{KGJ0}--\eqref{KGJz} can be generated from the known set of vectors by exploiting this freedom.

\nd \textit{(1) Linear sum of two series of generalized eigenvectors}

Given two series of generalized eigenvectors of the same order, $\Fm{z}$ and $G_N^{(z)}$ that satisfy \Eqs{KGJ0}--\eqref{KGJz}, any linear combination of them with constant coefficients,
\begin{equation}
   \label{H}
    H_N^{(z)}=c_1\Fm{z}+c_2G_N^{(z)}\,,
\end{equation}
 are also the generalized eigenvectors of $K_E$.
Note that the stationary states are always the same, $F_0^{(0)}=G_0^{(0)}=H_0^{(0)}$.
When we start with $\Fm{z}$ and $G_N^{(z)}$ that are adjoint-symmetric, cf.~the discussion at the beginning of  \ref{SecGJME}, we require $c_1$ and $c_2$ to be real so that $H_N^{(z)}$ is also adjoint-symmetric. We can then set $c_1+c_2=1$ to fix the overall constant of $H_N^{(z)}$.

\nd \textit{(2) Invariance of $K_E$ at the exceptional points}

It is convenient to adopt a matrix representation to discuss the invariance of $K_E$ under a similarity transformation.
We write $K_E$ in the subspace of generalized eigenvector of order $N$ \eqref{KGJ0}--\eqref{KGJz} as
\begin{equation}
   \label{KF}
    \KB_N \equiv\lam_N\left(\begin{array}{ccccccc}                      1 &0 &0 &\cdots&0&0&0 \\
                      1 &1 &0 &\cdots&0&0&0 \\
                      0 &1 &1 &\cdots&0&0&0 \\
                      \vdots&\vdots&\vdots&\ddots&\vdots&\vdots&\vdots \\
                      0&0&0&\cdots&1&1&0 \\
                      0&0&0&\cdots&0&1&1
\end{array}
\right)\,.
\end{equation}
 The generalized eigenvectors are
\begin{equation}
   \label{GJb}
    \FB_N^{(0)}\equiv\left(\begin{array}{c}                         0 \\
                         0 \\
                         0 \\
                         \vdots \\
                         0 \\
                         1
\end{array}
\right) \,,\quad
    \FB_N^{(1)}\equiv\left(\begin{array}{c}                         0 \\
                         0 \\
                         0 \\
                         \vdots \\
                         1 \\
                         0
\end{array}
\right) \,,\quad\cdots\,,\quad
    \FB_N^{(N-1)}\equiv\left(\begin{array}{c}                         0 \\
                         1 \\
                         0 \\
                         \vdots \\
                         0 \\
                         0
\end{array}
\right) \,,\quad
    \FB_N^{(N)}\equiv\left(\begin{array}{c}                         1 \\
                         0 \\
                         0 \\
                         \vdots \\
                         0 \\
                         0
\end{array}
\right)\,.
\end{equation}
 $\KB_N$ is then invariant under a similarity transformation
\begin{equation}
   \label{TK}
    \TB_N\cdot\KB_N\cdot\TB_N^{-1}=\KB_N\,,
\end{equation}
 where the matrix $\TB_N$ and its inverse are, respectively,
\begin{equation}
   \label{T}
    \TB_N \equiv\left(\begin{array}{cccccc}                      c_0&0&0&\cdots&0&0 \\
                      c_1&c_0&0&\cdots&0&0 \\
                      c_2&c_1&c_0&\cdots&0&0 \\
                      \vdots&\vdots&\vdots&\ddots&\vdots&\vdots \\
                       c_{N-1}&c_{N-2}&c_{N-3}&\cdots&c_0&0 \\
                      c_N&c_{N-1}&c_{N-2}&\cdots&c_1&c_0
\end{array}
\right)\,,
\qquad
    \TB_N^{-1}\equiv\left(\begin{array}{cccccc}                      d_0&0&0&\cdots&0&0 \\
                      d_1&d_0&0&\cdots&0&0 \\
                      d_2&d_1&d_0&\cdots&0&0 \\
                      \vdots&\vdots&\vdots&\ddots&\vdots&\vdots \\
                       d_{N-1}&d_{N-2}&d_{N-3}&\cdots&d_0&0 \\
                      d_N&d_{N-1}&d_{N-2}&\cdots&d_1&d_0
\end{array}
\right)\,.
\end{equation}
 The $c_i$ and $d_i$ are real coefficients and $c_0\neq0$.
They are related by
\begin{align}
   \label{dbar}
    d_0&=\frac{1}{c_0}\,,\\
    d_z&=-\frac{1}{c_0}\sum_{i=0}^{z-1} c_{z-i}\,d_i\,, \qquad z=1,2,3, \ldots,N.
\end{align}
 We note that the form \eqref{T} is preserved by $(\TB_N)^n$, where $n$ is a natural number.
The invariance in \Eq{TK} implies that
\begin{equation}
   \label{TF}
    \boldsymbol{G}_N^{(z)}\equiv\TB_N\cdot\FB_N^{(z)}\,, \qquad     \boldsymbol{H}_N^{(z)}\equiv\TB_N^{-1}\cdot\FB_N^{(z)}
\end{equation}
 are also generalized eigenvectors of $\KB_N$ satisfying \Eqs{KGJ0}--\eqref{KGJz}.
The generalized eigenvectors generated from $\Fm{z}$ are
\begin{equation}
\label{GJG}
\G{N}{z}=\sum_{i=0}^{z} c_i\F{N}{z-i}\,,\qquad \H{N}{z}=\sum_{i=0}^{z} d_i\F{N}{z-i} \qquad z=0,1,2, \ldots,N\,,
\end{equation}
 $c_0$ fixes the overall constant in the series of the generalized eigenvectors mentioned at the beginning of  \ref{SecFreedomGJ}.
Once the overall constant of $\F{N}{0}$ is fixed, the overall constant of other generalized eigenvectors in the series then follows.

As an illustration of this freedom, we list a few interesting cases for the subspace $N=3$.
\begin{enumerate}
\item If we set $c_i=1$ for all $i$, then $d_0=1, d_1=-1$, and $d_2=d_3=0$.
The following series of $\G{N}{z}$ and $\H{N}{z}$ are produced,
\begin{subequations}
\begin{align}
   \label{GJG1}
    \G{3}{0}&=\F{3}{0}\,,\qquad\qquad\qquad\qquad\quad     \H{3}{0}=\F{3}{0}\,,\\
    \G{3}{1}&=\F{3}{1}+\F{3}{0}\,,\qquad\qquad\qquad\,\,      \H{3}{1}=\F{3}{1}-\F{3}{0}\,,\\
    \G{3}{2}&=\F{3}{2}+\F{3}{1}+\F{3}{0}\,,\qquad\quad\,\,\,    \H{3}{2}=\F{3}{2}-\F{3}{1}\,.
\end{align}
\end{subequations}
\item Setting $c_0=1$, $c_1=c_3=0$, and $c_2=c$, we get
\begin{subequations}
\begin{align}
   \label{GJG2}
    \G{3}{0}&=\F{3}{0}\,,\qquad\qquad\qquad     \H{3}{0}=\F{3}{0}\,,\\
    \G{3}{1}&=\F{3}{1}\,,\qquad\qquad\qquad     \H{3}{1}=\F{3}{1}\,,\\
    \G{3}{2}&=\F{3}{2}+c\F{3}{0}\,,\qquad\quad    \H{3}{2}=\F{3}{2}-c\F{3}{0}\,,\\
    \G{3}{3}&=\F{3}{3}+c\F{3}{1}\,,\qquad\quad     \H{3}{3}=\F{3}{3}-c\F{3}{1}\,.
\end{align}
\end{subequations}
Notice that in this series, the even-order ($\F{3}{0}, \F{3}{2}$) and the odd-order ($\F{3}{1}, \F{3}{3}$) of generalized eigenvectors are mixed in separate subspace.
\item Setting $c_0=1$, $c_1=c_2=0$ and $c_3=d$, we can add arbitrary copies of $\F{3}{0}$ to the highest order vector in a series. Then only the generalized eigenvector of highest order is modified,
\begin{subequations}
\begin{align}
    \G{3}{z}&=\F{3}{z}\,, \qquad\qquad\qquad \H{3}{z}=\F{3}{z}\,, \qquad\qquad\qquad z=0,1,2\,,\\
    \G{3}{3}&=\F{3}{3}+d\F{3}{0}\,,\qquad\quad  \H{3}{3}=\F{3}{3}-d\F{3}{0}\,.  \label{Fmm}
\end{align}
\end{subequations}
\end{enumerate}

We can also apply the freedom to the generalized eigenvectors $\Phib{N}{z}$ and $\Phit{N}{z}$ listed in \ref{TeXFolio:tab1}  and \ref{TeXFolio:tab2},  respectively.
As shown in \ref{AppGJCLdiapar}, different choices of coefficients produce different series of generalized eigenvectors $\Psib{N}{z}$ for the Liouvillian of the CL equation, $\Kb_E$.
A few of them are listed in \ref{TeXFolio:tab3}.  In fact, the $\Phib{N}{z}$ and $\Psib{N}{z}$ are connected by a similarity transformation $\bar{\boldsymbol{\Psi}}_N=\bar{\TB}_N\cdot\bar{\boldsymbol{\Phi}}_N$, where for the first few subspaces $N=1,2,3$, they are
\begin{equation}
\label{PhibS}
\bar{\TB}_1=\left(\begin{array}{cc}               1 & 0 \\
               -1 & 1
\end{array}
           \right)\,,\qquad
    \bar{\TB}_2=\left(\begin{array}{ccc}               1 & 0 & 0 \\
               -2 & 1 & 0 \\
               2 & -2 & 1
\end{array}
           \right)\,,\qquad
    \bar{\TB}_3=\left(\begin{array}{cccc}               1 & 0 & 0 &0 \\
               -3 & 1 & 0 &0 \\
               \frac{9}{2} & -3 & 1 &0 \\
               -\frac{9}{2} & \frac{9}{2} & -3 & 1
\end{array}
           \right)\,,
\end{equation}
respectively.
\begin{table}[t]    \label{tabPibC2}
\begin{center}
\begin{tabular}{c|cccccccc}
   $z$ &\quad  &$N=0$ &\quad &$N=1$  &\quad &$N=2$   &\quad &$N=3$ \\
   \hline\\
   0 &\quad  &1 &\quad &$\Qb-i\rb$ &\quad &$(\Qb-i\rb)^2-\frac{1}{2}$ &\quad &$(\Qb-i\rb)^3-\frac{3}{2}(\Qb-i\rb)$\\\\
   1 &\quad  && &$-\Qb$    &\quad  &$-2\Qb(\Qb-i\rb)+1$    &\quad  &$-3(\Qb-i\rb)^3-3i\rb(\Qb-i\rb)^2+\frac{9}{2}(\Qb-i\rb)
    +\frac{3}{2}i\rb$  \\\\
   2 &\quad  &&  &\quad &    &$2\Qb^2-1$  &\quad &$\frac{9}{2}(\Qb-i\rb)\left(\Qb^2-\frac{1}{2}\right)-\frac{9}{2}\Qb$ \\\\
   3 &\quad  &&  &\quad  & &  &\quad  &$-\frac{9}{2}\Qb\left(\Qb^2-\frac{3}{2}\right)$
\end{tabular}
\end{center}
\caption{The first few series of $\Psib{N}{z}$.}
\end{table}

An alternate series of generalized eigenvectors of the Liouvillian of the modified KL equation, $\Psit{N}{z}$ using a different choice of coefficients chosen in \ref{AppGJKL}, is listed in \ref{TeXFolio:tab4}.
\begin{table}[t]    \label{tabPitD2}
\begin{center}
\begin{tabular}{c|cccccccccc}
   $z$ &\quad  &$N=0$ &\quad  &$N=1$  &\quad &$N=2$   &\quad &$N=3$ \\
   \hline\\
   0 &\quad  &1 &\quad &$i\wt\rt$ &\quad &$(i\wt\rt)^2$ &\quad &$(i\wt\rb)^3$ \\\\
   1 &\quad  &&\quad &$\Qt/2+i\wt\rt$ &\quad &$2i\wt\rt\left(\Qt/2+i\wt\rt\right)$ &\quad
   &$3(i\wt\rt)^2\left(\Qt/2+i\wt\rt\right)$ \\\\
   2 &\quad  &&\quad  & &\quad &$2\left(\Qt/2+i\wt\rt\right)^2-\frac{1}{4}$
     &\quad &$\frac{9}{2}(i\wt\rt)\left[\left(\Qt/2+i\wt\rt\right)^2-\frac{1}{8}\right]$\\\\
   3 &\quad  & &\quad & &\quad & &\quad   &$\frac{9}{2}\left(\Qt/2+i\wt\rt\right)
        \left[\left(\Qt/2+i\wt\rt\right)^2-\frac{3}{8}\right]$
\end{tabular}
\end{center}
\caption{The first few series of $\Psit{N}{z}$.}
\end{table}
It happens that the $\Phit{N}{z}$ and $\Psit{N}{z}$ are related by the inverse of the similarity transformation in \Eq{PhibS}, $\tilde{\boldsymbol{\Psi}}_N=\tilde{\TB}_N\cdot\tilde{\boldsymbol{\Phi}}_N$, where $\tilde{\TB}_N=\bar{\TB}_N^{-1}$.
The first few $\tilde{\TB}_N$ are
\begin{equation}
\label{Ttinv}
\tilde{\TB}_1=\left(\begin{array}{cc}                       1 & 0  \\
                       1 & 1
\end{array}
                   \right)\,, \qquad
    \tilde{\TB}_2=\left(\begin{array}{ccc}                       1 & 0 & 0 \\
                       2 & 1 & 0 \\
                       2 & 2 & 1
\end{array}
                   \right)\,, \qquad
    \tilde{\TB}_3=\left(\begin{array}{cccc}                       1 & 0 & 0 & 0 \\
                       3 & 1 & 0 & 0 \\
                       \frac{9}{2} & 3 & 1 & 0 \\
                       \frac{9}{2} &\frac{9}{2} &3 &1
\end{array}
                   \right)\,.
\end{equation}

\section{Evolution of quantum state at the exceptional points}
\label{SecEvoEP}
\subsection{Deviation from purely exponential decay}
\label{SecNonExpDec}

We know that as friction increases, the oscillatory motion of a classical
harmonic oscillator gradually vanishes, i.e.,~its underdamped motion goes into
critical damping and then followed by overdamping~\cite{Symon,Heiss16}.
At critical damping, the oscillator relaxes to its equilibrium position asymptotically with a time dependence of $e^{-\gam t/2}$ and $te^{-\gam t/2}$.
The generalized eigenvectors exhibit a similar time dependence at the exceptional point.
For example, this behaviour was also observed in Ref.~\cite{Bhamathi96} for
second-order resonance pole in Friedrichs model in which $te^{-\gam t}$ was
called a secular term.
Similar behaviours are also obtained in collective spin models, such as in
Refs.~\cite{Claeys22,Rubio-Garcia22}.

The solutions to a non-Hermitian eigenvalue problem can be worked out using the
method discussed in Ref.~\cite{Weintraub09}.
Let $\phi_N$ be an initial state consisting of the combination of generalized eigenvectors from the $N$-subspace with real coefficients $\phi_N=c_0 \Fm{0} +c_1\Fm{1}+\cdots+c_N\Fm{N}$. In matrix form, it is
\begin{equation}
   \label{Kphi}
    \phiB_N(0)=\left(\begin{array}{c}                         c_N \\
                         c_{N-1} \\
                         \vdots \\
                         c_1 \\
                         c_0
\end{array}
\right) \,.
\end{equation}
The solution to the matrix equation $\frac{\d}{\d t}\phiB_N=-\KB_N\cdot\phiB_N$ is~\cite{Weintraub09}
\begin{equation}
\label{phiBN}
\phiB_N(t)=e^{-\lam_N t}
    \left(\begin{array}{cccccc}                         1  & \\
                         -\lam_N t \\
                         \vdots& &\ddots\\
                        \displaystyle \frac{(-\lam_N t)^{N-2}}{(N-2)!}&\cdots &-\lam_N t &1 &0&0 \\
                         \displaystyle\frac{(-\lam_N t)^{N-1}}{(N-1)!}&\cdots &\displaystyle\frac{(\lam_N t)^2}{2!}&-\lam_N t &1&0 \\
                         \displaystyle\frac{(-\lam_N t)^N}{N!} &\cdots &\displaystyle-\frac{(\lam_N t)^3}{3!} &\displaystyle\frac{(\lam_N t)^2}{2!} &-\lam_N t &1
\end{array}
\right)\cdot \phiB_N(0)\,.
\end{equation}
For instance, in the $N=2$ subspace of the oscillator model we consider, where $\lam_2=\gam$, a state that starts initially as $\phi_2(0)=c_0F_2^{(0)}+c_1F_2^{(1)}+c_2F_2^{(2)}$ evolves into
\begin{subequations}
\begin{align}
   \label{nonExp2}
    \phi_2(t)&=e^{-\gam t} \left[
    \left(c_0-\gam t c_1 +\frac{\gam^2t^2}{2}c_2\right) F_2^{(0)}
    +(c_1-\gam tc_2)F_2^{(1)}
    +c_2F_2^{(2)} \right]\\
        &=e^{-\gam t}\phi_2(0)-\gam te^{-\gam t} \left(c_1F_2^{(0)}+c_2F_2^{(1)}\right)
                +\frac{(\gam t)^2}{2}e^{-\gam t}c_2F_2^{(0)}\,.
\end{align}
\end{subequations}
We find that deviations from the exponential decay of the initial state reveal themselves through polynomial terms in $t^n$.

\subsection{Relaxation of an excited state under {Caldeira--Leggett} ({ {CL}}) equation}
\label{SecRelaxCL}

\begin{figure}[t]
\centering
\includegraphics[width=6in,trim= 3cm 12cm 5.5cm 13cm]{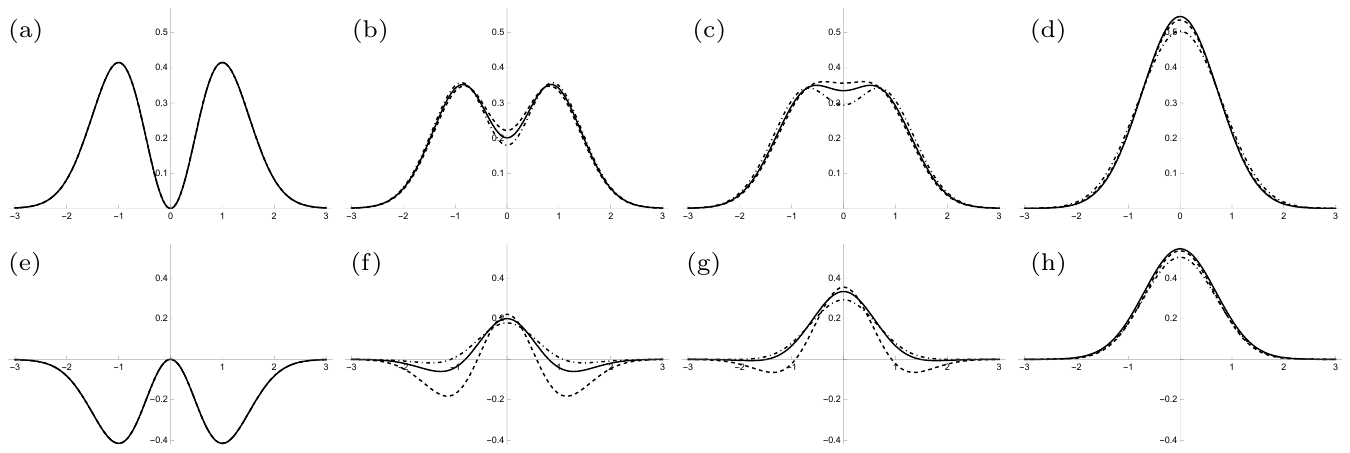}
\caption{Top and bottom row show the evolution of the probability distributions of $\bar{\Xi}_U(t), \bar{\Xi}_E(t)$ and $\bar{\Xi}_O(t)$ along the $\Qb$-axis with $\rb=0$, and their coherence components along the $\rb$-axis with $\Qb=0$, respectively, for three different damping at different time,  (a) and (e) at $\om_0t=0$, (b) and  (f) at $\om_0t=0.5$, (c) and (g) at $\om_0t=1$, (d) and (h) at $\om_0t=3$. Dashed, solid and dot-dashed lines denote underdamping ($\gam=\om_0$), critical damping ($\gam=2\om_0$), which corresponds to the exceptional point, and overdamping ($\gam=3\om_0$).\label{fig2}}
\end{figure}

In this section we follow the relaxation of the first excited state of a harmonic oscillator to the stationary state under the CL equation.
The first excited state has the wave function $\phi_1(x)=(2/\sqrt{\pi})^{1/2}x\exp(-x^2/2)$ in dimensionless
position coordinate.  Its density function is $\Xi(x,\xt)=\phi_1^{*}(x)\phi_1(\xt)$. It decomposes into
the eigenfunction of the KL equation $f_{mn}^\pm(Q,r)$ as (with
$b=1/2$)~\cite{Tay04}
\begin{equation}
    \Xi(Q,r)=f_{00}(Q,r)-2f_{10}(Q,r)
    =\frac{2}{\sqrt{\pi}}\left(Q^2-\frac{r^2}{4}\right)e^{-Q^2-r^2/4}\,.
\end{equation}
 This state has a two-peak probability distribution function $\Xi(x,0)$ along the diagonal ($r=0$), cf.~\Fig{fig2}(a). It has two troughs along the off-diagonal ($Q=0$), which is the coherence or interference component of the state, cf.~\Fig{fig2}(e). We can think of it as a superposition between two stationary wave packets giving rise to interference effect.
Because this state does not contain $f^\pm_{mn}$ component with non-zero $n$, the wave packet does not execute oscillating motion around the equilibrium position while it relaxes to the stationary state.
We consider three regions of damping below.

\begin{enumerate}[(a)]
\item Underdamped region, $\gam=\om_0$ (real $\om$).

Using the transformation $S_2$, we obtain the transformed state of $\Xi$ in the CL system. For simplicity, we take $\gam=\om_0$. Then substituting $\del_2=1-\gam/(2\om_0)=1/2$ into $\fb_{10}$ \eqref{Fb10} yields $S_2f_{10}(Q,r)=\fb_{10}(\Qb,\rb)=\Pib_{10}\fb_{00}$, where $\fb_{00}$ is the stationary state already given in \Eq{f00CL}, and the polynomial part is
\begin{equation}
    \Pib_{10}(\Qb,\rb)=\frac{2}{3}\left(1-2\Qb^2+2i\Qb\rb+2\rb^2\right)\,.
\end{equation}
Though a similarity transformation preserves the trace and the hermiticity (or adjoint-symmetry) of the density function $\Xi(Q,r)$~\cite{Tay17}, it does not guarantee the positivity of the transformed state
 $\fb_{00}-2\fb_{10}$.
We have to modify the coefficient of $\fb_{10}$ to $-3/2$ to ensure its positive definiteness,
\begin{equation}
   \label{Xib}
    \bar{\Xi}_U(0)=\fb_{00}-\frac{3}{2}\fb_{10}
    =\left(1-\frac{3}{2}\Pib_{10}\right)\fb_{00}\,.
\end{equation}
Notice that the trace or normalization is not altered by a change in the coefficient in front of the decaying state $\fb_{10}$ which has zero trace.
The time evolution is (with $\lam_{10}=\gam=\om_0$)
\begin{equation}
   \label{Xibt}
    \bar{\Xi}_U(t)=e^{-\Kb t}\bar{\Xi}(0)=\left(1-\frac{3}{2}e^{-\om_0 t}\Pib_{10}\right)\fb_{00}\,.
\end{equation}

\item Critically damped region, $\gam=2\om_0$ (at the exceptional point $\om=0$).

When approaching the exceptional points through $\gam\rightarrow2\om_0$, the dynamics is now determined by $\Kb_E$. Therefore, we need to write $\bar{\Xi}$ in the basis of the generalized eigenvectors $\Phib{N}{z}$ listed in
\ref{tabPibC2}. We find that
\begin{equation}
   \label{fb10Phi}
    \Pib_{10}=-\frac{4}{3}\left(\Phib{2}{0}-\frac{1}{2}\Phib{2}{1}+\frac{1}{2}\Phib{2}{2}\right)
    \,.
\end{equation}
 Consequently, we start with (cf.~\Eq{Xib})
\begin{equation}
   \label{XibEP}
    \bar{\Xi}_E(0)=\bar{\Xi}_U(0)
    =\left[1+2\left(\Phib{2}{0}-\frac{1}{2}\Phib{2}{1}+\frac{1}{2}\Phib{2}{2}\right)\right]
    \fb_{00}
\end{equation}
 as the initial state for time evolution at the exceptional point.
Using \Eq{nonExp2} and $\lam_2=\gam=2\om_0$, the evolution of $\bar{\Xi}_E$ is
\begin{equation}
   \label{XibEPt}
    \bar{\Xi}_E(t)=e^{-\Kb_E t}\bar{\Xi}_E(0)=\left(1+2e^{-2\om_0 t}\left[(1+\om_0t+\om_0^2t^2)\Phib{2}{0}-\frac{1}{2}(1+2\om_0t)\Phib{2}{1}
    +\frac{1}{2}\Phib{2}{2}\right]\right)\fb_{00}\,.
\end{equation}
\item Overdamped region, $\gam=3\om_0$ (imaginary $\om$).

When the relaxation rate is greater than $2\om_0$, the modified frequency of the oscillator becomes imaginary, $\om=i\sqrt{5}\om_0/2$. The oscillator now experiences overdamping. Substituting $\del_2=-1/2$ into the eigenfunctions in \Eqs{Fb10}--\eqref{Fb11pm}, we find that the decomposition of the initial condition involves also eigenfunctions from the $N=2$ subspace.
We will label the eigenfunctions of $\Kb_\CL$ in the overdamped region by $\bar{\Omega}^\pm_{mn}\fb_{00}$.
$\Pib_{10}(\Qb,\rb)$ can then be decomposed into them as
\begin{equation}
   \label{fbOver0}
    \Pib_{10}
    =\frac{1}{3}\Omb_{10}+\frac{2\sqrt{2}}{3}i\left(\Omb_{22}^+-\Omb_{22}^-\right)
    \,,
\end{equation}
 where
\begin{subequations}
\begin{align}
   \label{OmOver}
    \Omb_{10}&=-\frac{2}{5}(1-2\Qb^2+6i\Qb\rb+2\rb^2)\,,\\
    \Omb^\pm_{22}&=\mp\frac{i}{5\sqrt{2}}\left(3-6\Qb^2+8i\Qb\rb+6\rb^2\pm\sqrt{5}(1-2\Qb^2-2\rb^2)\right)
    \,.
\end{align}
\end{subequations}
The expressions of $\bar{\Omega}^\pm_{mn}$ can be obtained by setting $\del_2=-1/2$ in \Eqs{Fb10} and \eqref{Fb11pm}.
 The eigenvalues are $\lam_{10}=3\om_0$ and $\lam^\pm_{22}=(3\mp\sqrt{5})\om_0$.
Using the initial state
\begin{equation}
\label{XibOver0}
\bar{\Xi}_O(0)=\bar{\Xi}(0)=\left(1-\frac{1}{2}\Omb_{10}
    -\sqrt{2}i\left(\Omb_{22}^+-\Omb_{22}^-\right)\right)\fb_{00}
    \,,
\end{equation}
 the state evolves into
\begin{equation}
   \label{XibOvert}
    \bar{\Xi}_O(t)=\left(1-e^{-3\om_0 t}\left[\frac{1}{2}\Omb_{10}+\sqrt{2}i\left(e^{\sqrt{5}\om_0 t}\Omb_{22}^+-e^{-\sqrt{5}\om_0 t}\Omb_{22}^-\right)\right]\right)\fb_{00}
    \,.
\end{equation}
\end{enumerate}

In \Fig{fig2}, the evolution of $\bar{\Xi}_U(t)$, $\bar{\Xi}_E(t)$ and $\bar{\Xi}_O(t)$ are plotted on the same graph at different instants.
They are labelled by dashed lines, solid lines and dot-dashed lines, respectively.
It was reported in dissipative collective spin
models~\cite{Claeys22,Rubio-Garcia22} that there is a slow down of relaxation
at the exceptional point.
In the continuous variable model we consider, we find that there is indeed a slow down in the relaxation of the probability distribution function $\bar{\Xi}(Q,0)$ when it passes from the underdamping region across the critical damping (the exceptional point) region into the overdamping region, as can be observed from the plots on the first row of \Fig{fig2}.
The polynomial terms in $t$ \eqref{XibEPt} that was introduced at the exceptional point tend to slow down the relaxation at the initial stage of the evolution.

However, the relaxation of the coherence component $\bar{\Xi}(0,r)$ exhibits an opposite behaviour.
From the plots on the second row of \Fig{fig2}, we observe that at the initial stage, the decoherence is slowest in the underdamped oscillator.
This is because the coherence component selects the off-diagonal components ($Q=0$) from the density function.
For example, the $\Qb^2$ and $\rb^2$ terms in $\bar{\Xi}_U(\Qb,\rb)$ and $\bar{\Xi}_O(\Qb,\rb)$ have opposite signs. This results in different kinds of behavior in their relaxation.

There is also contribution from the exponential decay $\exp(-\gam t)$ that tends to speed up the relaxation when the damping rate $\gam$ increases from underdamping, across critical damping into overdamping region.
In the final stage of the evolution when all the fast decaying components become negligible, we are left with the slowest decaying component $\exp(-(3-\sqrt{5})\om_0 t)\Omb^+_{22}\approx \exp(-0.76\om_0 t)\Omb^+_{22}$ from the overdamped region \eqref{XibOvert}. That is why in the last figure on each row of \Fig{fig2} we see that the overdamped oscillator is left behind by the underdamped and critically damped oscillator which have almost arrived at their stationary state.

In the case of the modified KL equation, because the stationary state \eqref{f00KLmod} depends on the parameter $h_1$ which decides the transition to the exceptional point, the evolution of a state at and away from the exceptional point cannot be compared as easily as in the case of CL equation. Therefore, we do not discuss it here.
Nonetheless, when purely Gaussian states are considered, it is possible to
obtain the exact time evolution of the generic models using the results from
Ref.~\cite{Tay17b}.
We will present the results elsewhere.

\section{Conclusion}
\label{SecConclusion}

In this work we obtain the spectrum of the quantum Markovian Liouvillian for an oscillator in a generic environment at the exceptional points and clarify the structure of its eigenspace.
The exceptional points occur when the modified frequency of the oscillator vanishes, which corresponds to critical damping of the oscillator.
For a given natural frequency, the exceptional points lie on a circle in a two-parameter space.
All orders of degeneracy up to infinity occur simultaneously.
Consequently, the eigenspace also breaks into an infinite number of subspaces, each one has its own Jordan block structure.
The degeneracy allows freedom in the choice of generalized eigenvectors.
The freedom reveals itself as the invariance in the Jordan block structure of the Liouvillian under a similarity transformation whose form is obtained.
In the Kossakowski--Lindblad equation whose natural frequency is not modified, the eigenfunctions do not coalesce so that it remains a case of ordinary degeneracy.

\hfill

\noindent \textbf{Acknowledgments}

\hfill
This work is supported by the Ministry of Higher Education Malaysia (MOHE) under the Fundamental Research Grant Scheme (FRGS), Grant No.~FRGS/1/2020/STG07/UNIM/02/01.

\appendix

\section{Expressions of functions and operators in position coordinates}
\label{AppPi}

The explicit form of the polynomial $\Pi^\pm_{mn}(Q,r)$ in \Eq{fQr}
is~\cite{Tay04,Tay08}
\begin{equation}
   \label{PiQr}
    \Pi^\pm_{mn}(Q,r)
                =\sum_{\mu=0}^{m-n}\sum_{\nu=0}^{\mu}
                    \sum_{\sig=0}^{n} c^{\pm\mu\nu\sig}_{mn} \left(\frac{Q}{ \sqrt{2b}} \right)^{2(\mu-\nu)+n-\sig}
                     H_{2\nu+\sig} \left( \sqrt{ \frac{b}{2}} \,  r \right)\,,
\end{equation}
in which $H_{2\nu+\sig}$ denotes the Hermite polynomials. The coefficients are
\begin{equation}
\label{cmnpm}
c^{\pm\mu\nu\sig}_{mn}= (\pm 1)^{n+\sig}   \frac{ (-1)^{\mu+\nu} }{ i^n 2^{2\nu+\sig} \mu !} \sqrt{\frac{(m-n)!}{m!} }  \left(
\begin{matrix}
 m \\n+\mu
\end{matrix}
 \right)
               \left(
\begin{matrix}
 \mu \\\nu
\end{matrix}
 \right)
               \left(
\begin{matrix}
 n \\\sig
\end{matrix}
 \right)\,.
\end{equation}
For example,
\begin{align}
    \Pi^\pm_{11}(Q,r)&=\mp i \left(\frac{Q}{\sqrt{2b}}\pm \sqrt{\frac{b}{2}}r\right)\,,\\
    \Pi_{10}(Q,r)&=\frac{1}{2}-\frac{Q^2}{2b}+\frac{b}{2}r^2\,,\\
    \Pi^\pm_{22}(Q,r)&=\frac{1}{\sqrt{2}}\left(
    \frac{1}{2}-\frac{Q^2}{2b}\mp Qr-\frac{b}{2}r^2\right)\,.
\end{align}

In terms of the annihilation and creation operators of harmonic oscillator as well as in the position coordinates, the seven operators in \Eq{Kgen} are
\begin{equation}
\label{iL0}
iL_0\rho=i\bigm[\adg a,\rho\bigm]\doteq \frac{i}{2}\left(-\frac{\d^2}{\d Q\d r}+Qr\right)\rho\,,
\end{equation}
\begin{equation}
\label{iM1}
iM_1\rho=\frac{i}{4}\bigl([\adg\adg,\rho]+[aa,\rho]\bigr)
    \doteq\frac{i}{2}\left(\frac{\d^2}{\d Q\d r}+Qr\right)\rho\,,
\end{equation}
\begin{equation}
\label{iM2}
iM_2\rho=\frac{1}{4}\bigl([\adg\adg,\rho]-[aa,\rho]\bigr)
    \doteq -\frac{1}{2}\left(\frac{\d}{\d Q}Q+r\frac{\d}{\d r}\right)\,,
\end{equation}
\begin{equation}
\label{O0}
(O_0-I/2)\rho=\frac{1}{4}\left(
    \bigl[\adg,\{a,\rho\}\bigr]-\bigl[a,\bigl\{\adg,\rho\bigr\}\bigr]\right)
    \doteq -\frac{1}{2}\left(\frac{\d}{\d
    Q}Q-r\frac{\d}{\d r}\right)\rho\,,
\end{equation}
\begin{equation}
\label{O+}
O_+\rho=-\frac{1}{2}\bigl[a,\bigl[\adg,\rho\bigr]\bigr]
    \doteq \frac{1}{4}\left(\frac{\d^2}{\d Q^2}-r^2\right)\rho\,,
\end{equation}
\begin{equation}
\label{L1+}
L_{1+}\rho= -\frac{1}{4}\left(\bigl[\adg,\bigl[\adg,\rho\bigr]\bigr]+\bigl[a,\bigl[a,\rho\bigr]\bigr]\right)
    \doteq-\frac{1}{4}\left(\frac{\d^2}{\d Q^2}+r^2\right)\rho\,,
\end{equation}
\begin{equation}
\label{L2+}
L_{2+}\rho= \frac{i}{4}\left(\bigl[\adg,\bigl[\adg,\rho\bigr]\bigr]-\bigl[a,\bigl[a,\rho\bigr]\bigr]\right)
    \doteq-\frac{i}{2}r\frac{\d}{\d Q}\rho\,,
\end{equation}
where we use $\doteq$ to denote the coordinate representation of an operator.

\section{Dominant term under similarity transformation}
\label{App1stterm}
\subsection{ {Caldeira--Leggett} ({ {CL}}) equation}
\label{AppEPdamp}

The Liouvillian of the CL equation is related to that of the KL equation by a similarity transformation $S_2$ \eqref{SCL} through
$K_\CL=S_2 K_\KL S_2^{-1}$~\cite{Tay17,Tay20}.
The parameters in $S_2$ are $\eta=-2b\tanh\phi$ and $\sinh\phi=-\gam/(2\om_2)$. The coefficients of the two Liouvillians are related by $\om_0=\om_2\cosh\phi$ and $\bb=b/\cosh\phi$.

Applying a similarity transformation $S_2$ \eqref{SCL} to the coordinates yields
\begin{align}
   \label{Qrbarhat}
  \hat{\Qb}&\equiv S_2\frac{Q}{\sqrt{2b}}S_2^{-1}= \frac{1}{\sqrt{\cosh \phi}}  \left[\cosh(\phi/2)\Qb+\sinh(\phi/2)\frac{i}{2}\frac{\d}{\d\rb}\right]
  +\frac{\sinh\phi}{\sqrt{\cosh\phi}}\left[\cosh(\phi/2)i\rb-\sinh(\phi/2)\frac{1}{2}\frac{\d}{\d \Qb}\right]\,,\\
 i\hat{\rb}&\equiv S_2\sqrt{\frac{b}{2}}ir S_2^{-1}= \sqrt{\cosh\phi}\left[\cosh(\phi/2)i\rb-\sinh(\phi/2)\frac{1}{2}\frac{\d}{\d\Qb}\right]\,,
\end{align}
where
\begin{align}
   \label{CLphi}
        \cosh\phi&=\frac{\om_0}{\om_2}\,, \qquad \cosh(\phi/2)=\frac{1}{2\sqrt{\om_2}}\left(\sqrt{\om_0-\gam/2}
        +\sqrt{\om_0+\gam/2}\right)\,,\\
        \sinh\phi&=-\frac{\gam}{2\om_2} \,, \quad \sinh(\phi/2)=\frac{1}{2\sqrt{\om_2}}\left(\sqrt{\om_0-\gam/2}
        -\sqrt{\om_0+\gam/2}\right)\,.
\end{align}
Using the stationary state $\fb_{00}$ \eqref{f00CL} of $\Kb_\CL$, we obtain
\begin{align}
   \label{Fb1pS}
  \bar{f}_{11}^\pm&=S_2 f_{11}^\pm=\mp i(\hat{\Qb}\pm\hat{\rb})\fb_{00}
  =\frac{1\mp i}{2\sqrt{\del_2}}(\Qb-i\rb)\fb_{00}
  -\frac{1\pm i}{2\sqrt{2-\del_2}}(\Qb+i\rb)\fb_{00}\,,\\
     \bar{f}_{10}&=S_2f_{10}=\left(\frac{1}{2}-\hat{\Qb}^2+\hat{\rb}^2\right)\fb_{00}
  =\frac{1-2\Qb^2+4i\Qb\rb(1-\del_2)+2\rb^2}{2\del_2(2-\del_2)}\fb_{00}\,,
    \label{Fb10}\\
    \fb^\pm_{22}&=S_2f^\pm_{22}=\frac{1}{2\sqrt{2}}
    \left(1-2\hat{\Qb}^2\mp4\hat{\Qb}\hat{\rb}-2\hat{\rb}^2\right)\fb_{00}\no
    &=\frac{1}{2\sqrt{2}\del_2(2-\del_2)}
    \left(\pm i\bigl[(1-\del_2)(1-2\Qb^2+2\rb^2)+4i\Qb\rb\bigr]
    +\sqrt{\del_2(2-\del_2)}(1-2\Qb^2-2\rb^2)\right)\fb_{00} \,, \label{Fb11pm}
\end{align}
where $\del_2\equiv1-\gam/(2\om_0)$ \eqref{CLdel}.
Note that $\Pib^\pm_{mn}$ can be extracted from $\fb^\pm_{mn}=\Pib^\pm_{mn}\fb_{00}$.
We further note that $\fb_{11}^\pm$ belong to the $N=1$ subspace, whereas $\fb_{10}$ and $\fb^\pm_{22}$ are in the $N=2$ subspace.

\subsection{Modified  {Kossakowski--Lindblad} ({ {KL}}) equation}
\label{AppEPKLmod}

The modified KL equation can be obtained from the KL equation via a similarity
transformation $S_1\text{\eqref{S1}},\Kt_\m=S_1 K_\KL S_1^{-1}$~\cite{Tay17,Tay20}. The parameters in $S_1$
are given by $\tanh\psi=-h_1/(2\om_0)$, $\eta_0=-\eta_2\gam/h_1$ and $\eta_2=-2\gam b h_1/(4\om_0\om_1)$, where $\om_1$ is
the modified frequency \eqref{w1}.
The coefficient in the equation is related by $\bb=b\bigl(\om_1^2+\gam^2/4\bigr)/(\om_0\om_1)$.

Applying a similarity transformation $S_1$ \eqref{S1} to the coordinates produces the following results,
\begin{align}
   \label{Qrtilhat}
  \hat{\Qt}&=S_1\frac{Q}{\sqrt{2b}} S_1^{-1}
        =\frac{1}{\om_1}\sqrt{\frac{\om_0+h_1/2}{\om_0(\om_1^2+\gam^2/4)}}
            \left[\left(\om_1^2+\frac{\gam^2}{4}\right)\Qt+ \frac{\gam^2}{4}\frac{1}{2} \frac{\d}{\d \Qt}+\frac{\gam}{2}\frac{h_1}{2}i\rt\right]\,,\\
  i\hat{\rt}&=S_1\sqrt{\frac{b}{2}}ir S_1^{-1} =\sqrt{\frac{\om_0(\om_0+h_1/2)}{\om_1^2+\gam^2/4}}i\rt\,.
\end{align}
Making use of the stationary state $\ft_{00}$ \eqref{f00KLmod} of $\Kt_\m$, we obtain
\begin{align}
   \label{Ft1pS}
  \ft_{11}^\pm&=S_1f_{11}^\pm=\mp i(\hat{\Qt}\pm\hat{\rt})\ft_{00}\no
            &=\frac{-\wt}{\sqrt{1+\del_1(2-\del_1)\wt^2}}
            \left[\sqrt{2-\del_1} i\rt
                \pm i \frac{\sqrt{\del_1}}{1+\del_1\wt^2}
                    \left(\Qt+i\wt\rt
                    +\del_1\left[(2-\del_1)\wt^2\Qt-i\wt\rt\right]
                    \right)\right]\ft_{00}\,,\\
  \ft_{10}&=S_1f_{10}=\left(\frac{1}{2}-\hat{\Qt}^2+\hat{\rt}^2\right)\ft_{00}\no
        &=\frac{\wt^2}{2\bigl(1+\del_1\wt^2\bigr)^2}
        \left(4\rt^2
        +\del_1\left[
        1-2\Qt^2-4\wt i\Qt\rt-2(1-\wt^2)\rt^2
        \right]
        +\del_1^2\left[\wt^2(1-4\Qt^2)+4\wt i\Qt\rt\right]
        +2\del_1^3 \wt^2\Qt^2\right)\ft_{00}\,,\\
  \ft_{22}^\pm&=S_1f^\pm_{22}=\frac{1}{2\sqrt{2}}
  \left(1-2\hat{\Qt}^2\mp4\hat{\Qt}\hat{\rt}-2\hat{\rt}^2\right)\ft_{00}\no
        &=\frac{1}{2\sqrt{2}}
        \biggm[
        \frac{-2(2-\del_1)\wt^2\rt^2}{1+\del_1(2-\del_1)\wt^2}
        +\frac{\del_1\wt^2}{1+\del_1\wt^2}\nonumber\\
        &\qquad\qquad-\frac{2\del_1\wt^2}{(1+\del_1\wt^2)^2}
        \left(
        \bigl[1+\del_1(2-\del_1)\bigr]\Qt^2
        +2\wt(1-\del_1)i\Qt\rt
        -\frac{\wt^2(1-\del_1)^2\rt^2}{1+\del_1(2-\del_1)\wt^2\bigm]}
        \right)\no
        &\qquad\qquad  \pm i\frac{\sqrt{\del_1(2-\del_1)}\wt^2}{1+\del_1\wt^2}
        \left(4i\Qt\rt-\frac{4(1-\del_1)\wt\rt^2}{1+\del_1(2-\del_1)\wt^2}
        \right)
        \biggm]\ft_{00}\,,
\end{align}
where $\wt\equiv2\om_0/\gam$ \eqref{wt} and $\del_1\equiv1-h_1/(2\om_0)$ \eqref{del1}.

\section{Generalized eigenvectors of  {Caldeira--Leggett} ({ {CL}}) equation}
\label{AppGJCL}

In this appendix, we will work out the first few series of the generalized eigenvectors of $\Kb_E$ \eqref{KLEPQr} for the CL equation using \Eqs{dia}--\eqref{par}, and the Jordan block structure of the reduced dynamics \Eqs{KGJ0}--\eqref{KGJz} at the exceptional points.

\subsection{First few series of generalized eigenvectors}
\label{AppGJCLdiapar}

Starting from \Eq{dia} with $\Phib{0}{0}=1$, in the $N=1$ subspace we have $\Phib{1}{0}=\u{1}{0}\Qb+\v{1}{0}i\rb$. From the discussion in  \ref{SecFreedomGJ}, once we choose the overall constant of the lowest order generalized eigenvector $\Fm{0}$, the overall constants of other vectors are also fixed. We fix the overall constant by choosing $\u{1}{0}=1$. Substituting $\Phib{1}{0}$ into \Eq{dia}, we solve the equation to get $\v{1}{0}=-1$. Therefore, $\Phib{1}{0}=\Qb-i\rb$.
We continue by using \Eq{par} $z=N=1$. Substituting $\Phib{1}{1}=\u{1}{1}\Qb+\v{1}{1}i\rb$ into \Eq{par} yield $\p{1}{1}+\q{1}{1}=-1$. We can now make different choices of $\p{1}{1}$ and $\q{1}{1}$. This is the freedom stemming from the discussion in  \ref{SecFreedomGJ}. For example, we could choose
\begin{subequations}
\begin{align}
   \label{Pib11a}
    \p{1}{1}&=0\,,\qquad \q{1}{1}=-1\,,\qquad \Phib{1}{1}=-i\rb\,,\\
    \p{1}{1}&=-1\,,\qquad \q{1}{1}=0\,,\qquad \bar{\Psi}_1^{(1)}=-\Qb\,, \label{Pib11b}
\end{align}
\end{subequations}
 or many others.
All this choices produce valid series of generalized eigenvectors.

In the $N=2$ subspace, we continue with $\Phib{2}{0}=(\u{2}{0}\Qb+\v{2}{0}i\rb)\Phib{1}{0}+\w{2}{0}\Phib{0}{0}$. Substituting it into \Eq{dia} we get the condition $\u{2}{0}+\v{2}{0}=0$ and $\w{2}{0}=(\v{2}{0}-1)/4$. We fix the overall constant by choosing $\u{2}{0}=1$. Then $\v{2}{0}=-1$ and $\w{2}{0}=-1/2$, leading to $\Phib{2}{0}=(\Qb-i\rb)^2-1/2$.
We continue with the choice \Eq{Pib11a}.
The next vector in this series is $\Phib{2}{1}=(\u{2}{1}\Qb+\v{2}{1}i\rb)\Phib{1}{1}$. Upon substituting this choice of $\Phib{2}{1}$ into \Eq{dia} and simplify the expression, we find that $\u{2}{1}=2$, $\v{2}{1}=-2$. This means $\Phib{2}{1}=-2i\rb(\Qb-i\rb)$.
The last vector in this order is $\Phib{2}{2}=(\p{2}{2}\Qb+\q{2}{2}i\rb)\Phib{1}{1}+\s{2}{2}\Phib{0}{0}$. Applying \Eq{KGJz} onto it, we obtain $\p{2}{2}=0$, $\q{2}{2}=-2$, $\s{2}{2}=-\p{2}{2}/4=0$. We then conclude that $\Phib{2}{2}=2(i\rb)^2$.

We continue in this way for $N=3$ and summarize the results in the
\ref{tabPibC1}.
We can continue in this manner to generate all the high order generalized eigenvectors. The results can be generalized to \Eqs{PiCLdia}--\eqref{PiCLpar}.

Other set of generalized eigenvectors can be generated using the freedom discussed in  \ref{SecFreedomGJ}. For example, if we instead continue with \Eq{Pib11b}, we will get the series of generalized eigenvectors of $\Kb_E$ summarized in
\ref{tabPibC2}.
However, for this series of generalized eigenvectors, additional terms, $\x{N}{z}\Phib{N}{z}$ and $\x{N}{z}\Phib{N}{z}$, should be added to \Eqs{dia} and \eqref{par}, respectively.

\subsection{Proof of the equality of diagonal and parallel representation of $\Phib{N}{z}$}
\label{AppGJCLequal}

In this section, we prove by induction that the diagonal representation \eqref{PiCLdia} and parallel representation \eqref{PiCLpar} of $\Phib{N}{z}$ are indeed identical in the overlapping region, i.e.,~when $z=1,2, \ldots,N-1$.

First, we check that \Eq{PiCLdia} and \eqref{PiCLpar} are indeed identical for $N=2$ and 3. Then assuming that the right hand side (RHS) of \Eqs{PiCLdia,PiCLpar} are equal for $\Phib{N}{z}$ in the subspace $N=m-1$ and $m$, we will show that they are also equal in the subspace $N=m+1$.

We start by simplifying the expression obtained from the RHS of the diagonal representation \eqref{PiCLdia} for $N=m+1$, and show that it equals \Eq{PiCLpar} for $N=m+1$. The RHS of \Eq{PiCLdia} for $N=m+1$ is
\begin{align}
\label{proofCLPi}
(\Qb-i\rb)\frac{\Phib{m}{z}}{m^z}-\frac{m-z}{2}\frac{\Phib{m-1}{z}}{(m-1)^z}&=(\Qb-i\rb)\frac{-i\rb}{z}\frac{\Phib{m-1}{z-1}}{(m-1)^{z-1}}
            -\frac{m-z}{2}\frac{-i\rb}{z}\frac{\Phib{m-2}{z-1}}{(m-2)^{z-1}}\no
        &=\frac{-i\rb}{z}\left((\Qb-i\rb)\frac{\Phib{m-1}{z-1}}{(m-1)^{z-1}}
            -\frac{(m-1)-(z-1)}{2}\frac{\Phib{m-2}{z-1}}{(m-2)^{z-1}}\right)\no
        &=\frac{-i\rb}{z}\frac{\Phib{m}{z-1}}{m^{z-1}}\,,\qquad z=1,2, \ldots,m\,,
\end{align}
 which is the RHS of \Eq{PiCLpar} for $N=m+1$.
In the first equality, we have made used of \Eq{PiCLpar} twice to replace $\Phib{m}{z}$ and $\Phib{m-1}{z}$, which are assumed to be correct and are valid for $z\neq0$.
To obtain the last equality, we use \Eq{PiCLdia} for $N=m$ and for $z-1$ to replace the expression in the bracket by $\Phib{m}{z-1}/m^{z-1}$. Lastly, \Eq{proofCLPi} is also valid for $z=m$ since by definition we set $\Phib{m-1}{m}=0$.
Therefore, \Eqs{PiCLdia} and \eqref{PiCLpar} are equal in the overlapping region.

\subsection{Proof that $\Phib{m}{z}$ are generalized eigenvectors of $\Kb_E$}
\label{AppGJCLproof}

We will show that $\Phib{N}{z}$ in \Eq{GJdecomp} satisfies \Eqs{KGJ0} and \eqref{KGJz}. We start with $z=0$. We know that $\Phib{N}{0}$ for $N=0,1,2, 3$ satisfy \Eqs{KGJ0}--\eqref{KGJz} by construction.
We prove the rest by induction. We first assume that \Eq{KGJ0}--\eqref{KGJz} are correct for arbitrary $N=m$, then we deduce that they are correct for $N=m+1$ too.
First, using \Eq{PiCLdia}, we have
\begin{equation}
\label{Pib0Qr}
\Fb{N}{0}=\Phib{N}{0}\Fb{0}{0}=(\Qb-i\rb)\F{N-1}{0}-\frac{N-1}{2} \Fb{N-2}{0}\,.
\end{equation}
Then, acting $\Kb_E$ \eqref{KLEPQr} on it, we obtain
\begin{equation}
\label{KPib0}
\Kb_E\Fb{m}{0}=(\Qb-i\rb)\Kb_E\Fb{m-1}{0}-\frac{m-1}{2}\Kb_E\Fb{m-2}{0}
        -\frac{\gam}{4}\left(\frac{\d\Fb{m-1}{0}}{\d\Qb}+i\frac{\d\Fb{m-1}{0}}{\d\rb}+4i\rb\F{m-1}{0}\right)
\end{equation}
Using $\Kb_E\Fb{m-1}{0}=\lam_{m-1}\Fb{m-1}{0}$, $\lam_{m-1}=(m-1)\gam/2$ and a similar expression for $N=m-2$ on the RHS of \Eq{KPib0}, we put the RHS of \Eq{KPib0} in the following form,
\begin{equation}
\Kb_E\Fb{m}{0}=\lam_m\left((\Qb-i\rb)\F{m-1}{0}-\frac{m-1}{2}\Fb{m-2}{0}\right)
        -\frac{\gam}{2}\left(\frac{1}{2}\frac{\d\Fb{m-1}{0}}{\d\Qb}
        +\frac{i}{2}\frac{\d\Fb{m-1}{0}}{\d\rb}+\Fb{m}{0}+2i\rb\F{m-1}{0}-\frac{m-1}{2}\Fb{m-2}{0} \right)\,.
\end{equation}
Notice that the first bracket equals $\F{m}{0}$.
Therefore, we are left to show that the rest of the terms in the second bracket vanishes. That is, defining
\begin{equation}
   \label{DN}
    D_N(\Qb,\rb)\equiv\frac{1}{2}\frac{\d\Fb{N-1}{0}}{\d\Qb}+\frac{i}{2}\frac{\d\Fb{N-1}{0}}{\d\rb}\,,
\end{equation}
we want to show that
\begin{equation}
\label{KPib0last}
D_N(\Qb,\rb)=-\Fb{N}{0}-2i\rb\Fb{N-1}{0}+\frac{N-1}{2}\Fb{N-2}{0}\,, \qquad N\geq2\,.
\end{equation}
Let us prove \Eq{KPib0last} by induction, i.e.,~we show that given \Eq{KPib0last} is true for $D_m$ and $D_{m-1}$, then it is also true for $D_{m+1}$.

We first verify that \Eq{KPib0last} is indeed true for $N=2,3$ by direct calculation.
Then we substitute the expression of $\Fb{m}{0}$ \eqref{Pib0Qr} into the RHS of \Eq{DN} for $N=m+1$. After carrying out differentiation, we simplify it to
\begin{equation}
\label{DN+1}
D_{m+1}(\Qb,\rb)=\frac{1}{2}\frac{\d\Fb{m}{0}}{\d\Qb}+\frac{i}{2}\frac{\d\Fb{m}{0}}{\d\rb}=\F{m-1}{0}+(\Qb-i\rb)D_m(\Qb,\rb)-\frac{m-1}{2}D_{m-1}(\Qb,\rb)\,.
\end{equation}
Assuming that \Eq{KPib0last} is true for $N=m-1$ and $m$, we substitute them into $D_m$ and $D_{m-1}$ in \Eq{DN+1}. After simplifying the expression we get
\begin{equation}
\label{DN+1b}
D_{m+1}(\Qb,\rb)=-\Fb{m+1}{0}-2i\rb\Fb{m}{0}+\frac{m}{2}\Fb{m-1}{0}\,,
\end{equation}
which is \Eq{KPib0last} for $N=m+1$. This proves that \Eq{PiCLdia} indeed satisfies \Eq{KGJ0}.

We then prove that \Eq{PiCLpar}, or $\Phib{N}{z}$ for $z\neq0$, satisfy \Eq{KGJz} by induction. We know that \Eq{PiCLpar} satisfy \Eq{KGJz} for $N=1,2$ by construction.
Then we assume that \Eq{KGJz} is true for $N=m-1$, and show that it is true for $N=m$ too.
We start with \Eq{PiCLpar} by multiplying it with $\Fb{0}{0}$ to get
\begin{equation}
\label{Fbmz}
\frac{\F{m}{z}}{m^z}=-\frac{i\rb}{z}\frac{\Fb{m-1}{z-1}}{(m-1)^{z-1}}\,.
\end{equation}
Then we act $\Kb_E$ \eqref{KLEPQr} on it to get
\begin{equation}
   \label{KPibz}
    \Kb_E\left(\frac{\Fb{m}{z}}{m^z}\right)
    =\frac{-i\rb}{z}\Kb_E\left(\frac{\Fb{m-1}{z-1}}{(m-1)^{z-1}}\right)
    -\gam\frac{i\rb}{z}\frac{\Fb{m-1}{z-1}}{(m-1)^{z-1}}
    -\frac{\gam}{4z}\frac{\d}{\d\Qb}\left(\frac{\Fb{m-1}{z-1}}{(m-1)^{z-1}}\right)\,.
\end{equation}
Let us calculate the first term on the RHS of \Eq{KPibz} using the  of the generalized eigenvalue equation \eqref{KGJz}, and then use \Eq{PiCLpar} to substitute $\Fb{m-1}{z-1}$ and $\Fb{m-1}{z-2}$ in the expression. The result is
\begin{equation}
   \label{KPibzb}
    \frac{-i\rb}{z}\Kb_E\left(\frac{\Fb{m-1}{z-1}}{(m-1)^{z-1}}\right)
    =\lam_{m-1}\frac{-i\rb}{z}\frac{\Fb{m-1}{z-1}}{(m-1)^{z-1}}
        +\lam_{m-1}\frac{-i\rb}{z}\frac{\Fb{m-1}{z-2}}{(m-1)^{z-2}}
     =\lam_m\left(\frac{\Fb{m}{z}}{m^z}+\frac{\Fb{m}{z-1}}{m^z}\right)
        -\frac{\gam}{2}\left(\frac{\Fb{m}{z}}{m^z}+\frac{1}{z}\frac{\Fb{m}{z-1}}{m^{z-1}}\right)\,,
\end{equation}
where we use \Eq{Fbmz} and $\lam_{m-1}=\lam_m-\gam/2$ to get the last equality.
Notice that the two terms in the first bracket on the RHS of the last equality equal $\Kb_E\bigl(\Fb{m}{z}/(\lam_N m^z)\bigr)$,
i.e., the left hand side (LHS) of \Eq{KPibz}.
Substituting \Eqs{KPibzb} back into \Eq{KPibz}, we are left to show that
\begin{equation}
   \label{KPibzd}
   \frac{1}{2z}\frac{\d}{\d\Qb}\left(\frac{\Fb{m-1}{z-1}}{(m-1)^{z-1}}\right)
    =-\frac{\Fb{m}{z}}{m^z}-\frac{1}{z}\frac{\Fb{m}{z-1}}{m^{z-1}}
        -\frac{2i\rb}{z}\frac{\Fb{m-1}{z-1}}{(m-1)^{z-1}}\,.
\end{equation}
 Denoting the LHS of \Eq{KPibzd} by
\begin{equation}
\label{DN'}
D'_{N,z}(\Qb,\rb)=\frac{1}{2z}\frac{\d}{\d\Qb}\left(\frac{\Fb{N-1}{z-1}}{(N-1)^{z-1}}\right)\,,
\end{equation}
we assume that if \Eq{KPibzd} were true for $N=m$ and $z\neq0$, then it is also true for $N=m+1$ for $z\neq0$.
We calculate $D'_{m+1,z}(\Qt,\rt)$ by substituting $\Fb{m}{z-1}$ using \Eq{PiCLpar} for $N=m$ and for $z\rightarrow z-1$ to get
\begin{equation}
\label{KPibze}
D'_{m+1,z}(\Qb,\rb)=\frac{1}{2z}\frac{\d}{\d\Qb}\left(\frac{\Fb{m}{z-1}}{m^{z-1}}\right)
    =\frac{-i\rb}{z}\frac{1}{2(z-1)}\frac{\d}{\d\Qb}
    \left(\frac{\Fb{m-1}{z-2}}{(m-1)^{z-2}}\right)\,.
\end{equation}
Then using \Eq{KPibzd} for $z\rightarrow z-1$ to substitute the term on the RHS of \Eq{KPibze}, we finally obtain
\begin{equation}
\label{D'last}
D'_{m+1,z}(\Qb,\rb)=\frac{-i\rb}{z}\left(-\frac{\Fb{m}{z-1}}{m^{z-1}}
            -\frac{1}{z-1}\frac{\Fb{m}{z-2}}{m^{z-2}}
        -\frac{2i\rb}{z-1}\frac{\Fb{m-1}{z-2}}{(m-1)^{z-2}}\right)
        =-\frac{\Fb{m+1}{z}}{(m+1)^z}-\frac{1}{z}\frac{\Fb{m+1}{z-1}}{(m+1)^{z-1}}
        -\frac{2i\rb}{z}\frac{\Fb{m}{z-1}}{m^{z-1}}
\end{equation}
where we repeatedly use \Eq{PiCLpar} to replace $\Fb{m}{z-1}, \Fb{m}{z-2}$ and $\Fb{m-1}{z-2}$ by $\Fb{m+1}{z}, \Fb{m+1}{z-1}$ and $\Fb{m}{z-1}$, respectively, to get the last equality.
Now notice that \Eq{D'last} is \Eq{KPibzd} for $N=m+1$ for $z\neq0$ as we set up to prove.
This completes the proof that \Eqs{PiCLdia}--\eqref{PiCLpar} satisfy \Eqs{KGJ0}--\eqref{KGJz} for $\Kb_E$.

\section{Generalized eigenvectors of modified {Kossakowski--Lindblad} ({ {KL}}) equation}
\label{AppGJKL}

In this appendix, we will work out the first few series of the generalized eigenvectors of the modified KL equation with the Liouvillian $\Kt_E$ \eqref{KLmEPQr} at the exceptional point, using the similar method in the last section.

We start with $\Phit{0}{0}=1$. From \Eq{dia}, the next order vector is $\Phit{1}{0}=\u{1}{0}\Qt+\v{1}{0}i\rt$. Substituting $\Phit{1}{0}$ into \Eq{KGJz}, we obtain the condition $\u{1}{0}=0$ only. There is no requirement on $\v{1}{0}$. We choose $\v{1}{0}=\wt$, where $\wt\equiv 2\om_0/\gam$ \eqref{wt}.
Other choices will give different set of generalized eigenvectors.
This ambiguity is attributed to the freedom discussed in  \ref{SecFreedomGJ}.
Then $\Phit{1}{0}=i\wt\rt$.
We then use \Eq{par} to get the next vector in this series $\Phit{1}{1}=\p{1}{1}\Qt+\q{1}{1}\rt$.
Acting $\Kt_E$ on $\Phit{1}{1}$, we obtain $\p{1}{1}=1/2$, with no constraint on $\q{1}{1}$. The simplest choice is $\q{1}{1}=0$. As a result, $\Phit{1}{1}=\frac{\Qt}{2}$.

Continuing on to $N=2$, applying \Eq{KGJ0} on $\Phit{2}{0}=(\u{2}{0}\Qt+\v{2}{0}i\rt)\Phit{1}{0}+\w{2}{0}\Phit{0}{0}$ yield $\u{2}{0}=0$ and $\w{2}{0}=0$. Choosing the free coefficient $\v{2}{0}=\wt$, we get
$\Phit{2}{0}=(i\wt\rt)^2$.
The next generalized eigenvector is $\Phit{2}{1}=(\u{2}{1}\Qt+\v{2}{1}i\rt)\Phit{1}{1}$.
Substituting in $\Phit{1}{1}$, \Eq{KGJz} then produces the condition $\u{2}{1}=0$ and $\v{2}{1}=2\wt$ to yield $\Phit{2}{1}=i\wt\rt\Qt$.
The last generalized eigenvector in this series is $\Phit{2}{2}=(\p{2}{2}\Qt+\q{2}{2}i\rt)\Phit{1}{1}+\s{2}{2}\Phit{0}{0}$, see \Eq{par}. Applying \Eq{KGJz} for $z=2$ then gives $\p{2}{2}=1$, $\q{2}{2}=0$ and $\s{2}{2}=-\p{2}{2}/4=-1/4$. Hence, $\Phit{2}{2}=\frac{1}{2}\Qt^2-\frac{1}{4}$.
Continuing in this way, we obtain the generalized eigenvectors for $N=3$.
We can continue in this way to generate higher order generalized eigenvectors.
The results can be generalized to \Eqs{PiKLmdia}--\eqref{PiKLmpar}.
The first few series of the generalized eigenvectors are summarized in
\ref{tabKlmod}.

The proof that \Eqs{PiKLmdia} and \eqref{PiKLmpar} are equal in the overlapping region and that they lead to generalized eigenvectors \eqref{GJdecomp} of $\Kt_E$ that satisfy \Eqs{KGJ0}--\eqref{KGJz} can be proved by induction, similar to the corresponding proof for the generalized eigenvectors of $\Kb_E$ in \ref{AppGJCL}. We omit the proof here.

If we start with a different choice of the coefficients, such as $\q{1}{1}=\wt$, we will get a different series of generalized eigenvectors.
However, we note that additional terms, such as $\x{N}{z}\Phib{N}{z}$ and $\x{N}{z}\Phib{N}{z}$, should be added to the RHS of \Eqs{dia} and \eqref{par}, respectively, for this series of generalized eigenvectors.
The results are listed in
\ref{tabPitD2}.

\section{ {Hu--Paz--Zhang} equation}
\label{AppHPZ}

The Liouvillian of the Markovian limit of the HPZ equation~\cite{Tay06} differs
from that of $\Kb_\CL$ by an extra term, $K'_\HPZ=\Kb_\CL-dL_{2+}$.
It can be obtained from $\Kb_\CL$ by a similarity transformation $K'_\HPZ=S_3\Kb_\CL S_3^{-1}$, where
\begin{equation}
\label{SHPZ}
S_3=e^{\zeta L_{1+}}\,.
\end{equation}
In terms of the following coordinates,
\begin{equation}
   \label{HPZQr}
    Q_+\equiv \frac{Q}{\sqrt{2b_+}}\,,\qquad  r_-\equiv \sqrt{\frac{b_-}{2}} r\,,
\end{equation}
where
\begin{equation}
  \label{b+-}
    b_+\equiv b_\HPZ+\frac{d}{2\om_0} =\bb+ \frac{d}{4\om_0}\,, \qquad
    b_-\equiv b_\HPZ=\bb- \frac{d}{4\om_0}\,.
\end{equation}
The stationary state is
\begin{equation}
   \label{feqHPZ}
       f'_{00}(Q_+,r_-)\equiv S_3\fb_{00}(\Qb,\rb)=\frac{1}{\sqrt{\pi}}e^{-Q_+^2-r_-^2}\,.
\end{equation}
We refer the readers to Ref.~\cite{Tay20} for the expressions of $\Phi^{\prime \pm}_{mn}$,
cf. \Eq{GJdecomp}.

\subsection{Hu--Paz--Zhang ({ {HPZ}}) equation}
\label{SecEPHPZ}

In the position coordinates \eqref{HPZQr}, the Markovian limit of the HPZ equation has the Liouvillian
\begin{equation}
\label{KHPZQr}
K'_\HPZ(Q_+,r_-)=i\om_0\left(-\frac{1}{2}\sqrt{\frac{b_-}{b_+}}\frac{\d^2}{\d Q_+\d r_-}
        +2\sqrt{\frac{b_+}{b_-}} Q_+r_-\right)
                    +\gam\rb\frac{\d}{\d \rb}+2\gam\rb^2+i\om_0\frac{b_+-b_-}{\sqrt{b_+b_-}}r_-\frac{\d}{\d Q_+}\,.
\end{equation}
The transformed coordinate operators are
\begin{subequations}
\begin{align}
   \label{Qhat+}
    \hat{Q}_+&\equiv S_3\Qb S_3^{-1}
            =\frac{1}{\sqrt{2b_+(b_++b_-)}}
            \left[2b_+Q_++\frac{1}{2}(b_+-b_-)\frac{\d}{\d Q_+}\right] \,, \\
    \hat{r}_-&\equiv S_3\rb S_3^{-1}=\sqrt{\frac{b_++b_-}{2b_-}}r_-\,.
\end{align}
\end{subequations}

$K'_\HPZ$ has the same exceptional point as $\Kb_\CL$, that is at $\gam=2\om_0$.
We replace this into \Eq{KHPZQr} to get the Liouvillian of HPZ equation at the exceptional point $K'_E$.
We can obtain the generalized eigenvectors of $K'_E$ directly from those of $\Kb_E$ by similarity transformation.
They are
\begin{equation}
   \label{GJHPZ}
    F_N^{\prime(z)}\equiv S_3 \Fb{N}{z}\,,
\end{equation}
with the stationary state $F_0^{\prime(0)}=f^\prime_{00}$ \eqref{feqHPZ}.
The higher order vectors can be generated using
\begin{align}
   \label{S3QrF00}
    S_3\Qb\Fb{0}{0}&=\hat{Q}_+F_0^{\prime(0)}
        =\sqrt{\frac{b_++b_-}{2b_+}}Q_+F_0^{\prime(0)}\,,\\
    S_3(i\rb)\Fb{0}{0}&=i\hat{r}_-F_0^{\prime(0)}
        =\sqrt{\frac{b_++b_-}{2b_-}}ir_-F_0^{\prime(0)}\,.
\end{align}
Applying $S_3$ to \Eqs{PiCLdia} and \eqref{PiCLpar}, the results are
\begin{subequations}
\begin{align}
   \label{PiHPZdia}
    \text{Diagonal}:\qquad &\frac{\Phi^{\prime(z)}_N}{N^z}
    =\sqrt{\frac{b_++b_-}{2b_+}}\left(Q_+-\sqrt{\frac{b_+}{b_-}}ir_-\right)\frac{\Phi^{\prime(z)}_{N-1}}{(N-1)^z}
    -\frac{N-1-z}{2}\frac{\Phi^{\prime(z)}_{N-2}}{(N-2)^z}\,,\qquad z\neq N\,,\\
    \text{Parallel}:\qquad &\frac{\Phi^{\prime(z)}_N}{N^z}=\sqrt{\frac{b_++b_-}{2b_-}}\frac{1}{z}(-ir_-)
    \frac{\Phi^{\prime(z-1)}_{N-1}}{(N-1)^{z-1}}\,,\qquad\qquad\qquad\qquad\qquad\qquad\qquad z\neq 0\,.  \label{PiHPZpar}
\end{align}
\end{subequations}
As usual, we set $\Phi^{\prime(z)}_N=0$, whenever $z< 0, z > N$ or $N<0$.
We can then construct the whole series of generalized eigenfunctions successively starting with $\Phi_0^{\prime(0)}=1$.

\providecommand{\noopsort}[1]{}\providecommand{\singleletter}[1]{#1}%

\end{document}